\documentclass[numbib]{imamat}
\usepackage{amsmath}
\usepackage{amssymb}
\usepackage{graphicx}
\usepackage[lofdepth,lotdepth]{subfig}

\begin{document}
\renewcommand{\baselinestretch}{1.1}
\newcommand{\dd}{d}
\newcommand{\In}{M}
\newcommand{\UU}{U}
\newcommand{\hh}{h}
\newcommand{\rb}{\rangle}
\newcommand{\lb}{\langle}
\newcommand{\la}{\lambda}
\def\cistar{\kern.2em\mbox{$\odot\kern-.67em\star\kern .4em$}}
\newcommand{\laL}{{\lambda \in \Lambda}}
\newcommand{\wS}{\widetilde S}
\newcommand{\om}{\omega}
\newcommand{\Real} {{\rm Real}}
\newcommand{\R} {{\mathbb R}}
\newcommand{\M} {{P}}
\newcommand{\E} {{\mathbb E}}
\newcommand{\N} {{\mathbb N}}
\newcommand {\h} {{\mathfrak h}}
\newcommand{\Z} {{\mathbb Z}}
\newcommand{\C} {{\mathbb C}}
\newcommand{\HH} {{\cal H}}
\newcommand{\Ld} {{\bf L^2}}
\newcommand{\LD} {{\bf L^2}({\mathbb R}^{d})}
\newcommand{\Lu} {{\bf L^1}}
\newcommand{\Ga} {{\Gamma}}
\title{Phase Harmonic Correlations
and Convolutional Neural Networks}
\shorttitle{Phase Harmonic Correlations and Convolutional Neural Networks}
\shortauthorlist{St\'ephane Mallat, Sixin Zhang and Gaspar Rochette}

\author{
\name{St\'ephane Mallat}
\address{Coll\`ege de France, Paris\\
DI, \'Ecole Normale Sup\'erieure, PSL Researh University, Paris\\
CCM, Flatiron Institute, New York}
\name{Sixin Zhang}
\address{DI, \'Ecole Normale Sup\'erieure, PSL Researh University, Paris\\
Center for Data Science, Peking University, Beijing, China}
\and
\name{Gaspar Rochette}
\address{DI, \'Ecole Normale Sup\'erieure, PSL Researh University, Paris}
}

\maketitle

\begin{abstract}
{A major issue in harmonic analysis is to capture the phase dependence of
frequency representations, which carries important signal properties.
It seems that convolutional neural networks have found a way.
Over time-series and images,
convolutional networks often learn a first layer of filters which 
are well localized in the frequency domain, with different
phases. We show that a rectifier then acts as a filter on the phase of 
the resulting coefficients. It computes signal descriptors which
are local in space, frequency and phase. The non-linear phase filter
becomes a multiplicative operator over
phase harmonics computed with a Fourier transform along the phase. 
We prove that
it defines a bi-Lipschitz and invertible representation.
The correlations of phase harmonics coefficients characterise
coherent structures from their phase dependence across frequencies.
For wavelet filters, we show numerically that signals having
sparse wavelet coefficients can be recovered from 
few phase harmonic correlations, which provide a compressive 
representation.}
{Neural networks, harmonics, phase, wavelets}\\ 
2000 Math Subject Classification: 42C40, 62M45
\end{abstract}

\section{Introduction}

Convolutional
neural networks capture highly complex properties
of one, two and three-dimensional signals, leading to remarkable 
applications for classification, regression and generation \citep{CNNnature}.
For image and time-series applications, the learned filters in the first layer
are usually well localized both in space or time, and along frequencies,
with different 
phases \citep{Imagenet,Tasnet}. They resemble wavelets \citep{mallatReview}.
These filters provide a representation of an input signal $x(u)$ which is 
local along $u$ and along frequencies. However, 
rectifiers applied to first layer convolutional network coefficients
also perform a filtering along phases, which creates multiple harmonics.
We show that rectified coefficients define a local 
space-frequency-phase representation where local signal properties 
can be characterised by linear correlations across phases and frequencies.

The mathematical properties of rectifiers 
are more easily analyzed by extending first layer real network
coefficients into complex values. This is done with 
a complex analytic extension of the network filters.
A central result shows that
rectifiers act as a filter along the analytic phase.
By applying a Fourier transform along the phase, 
the phase filtering becomes a multiplicative
operator over phase harmonics. 
Phase harmonics can be interpreted as non-linear frequency transpositions.
For appropriate network filters, they
define an invertible bi-Lipschitz signal representation.
We shall study the particular case where network 
filters are wavelets, and thus
define a complete and stable multiscale representation. 

A major issue in harmonic analysis is to capture the dependence of
phases across frequencies, which produce 
 ``coherent structures'' such as isolated singularities. 
The phase is often considered
as an annoying variable which yields decorrelated coefficients at different
frequencies because of cancellation effects.
Moreover, the phase is unstable at locations
the modulus vanishes. Most recent harmonic analysis research
has concentrated on the modulus of wavelet coefficients,
which is sufficient to specify uniform or pointwise regularity of functions in
Sobolev, H\"{o}lder or Besov spaces 
\citep{jaffard}. However, scrambling the phase by modifying the sign 
of wavelet coefficients in a basis destroys the geometric
structures of signals and images, 
which shows that phase dependencies are crucial. The importance of the
wavelet phase to capture coherent signal structures
was emphasized by the pioneer work in \citep{Grossmann}.
It was also shown that correlating the phase of wavelet coefficients 
across frequencies improves the synthesis of image textures
\citep{Portilla}.

The main contribution of this paper is to define a stable
local phase harmonic representation, where 
phase dependencies can be captured by linear autocorrelation matrices.
Section \ref{invarian} analyzes their mathematical properties. The use
of autocorrelation matrices over deep convolutional network coefficients
was introduced in \citep{Bethdge1}, to generate
image textures which are perceptually similar to an original one.
These results are simplified in \citep{Bethdge2}, by
synthesizing textures from correlations of coefficients 
in a first convolutional layer, computed with local filters.
We shall concentrate on recovery of signal with small $\bf L^2$
approximation errors as opposed
to signals having similar perceptual properties.

Section \ref{reconsec} shows that fixed sets of
wavelet harmonic correlations can be sufficient to
recover close approximations of signals
having sparse wavelet coefficients.
These properties are similar to
compressive sensing algorithms \citep{candes},
where linear sensing measurements are replaced by 
non-linear harmonic autocorrelations. Numerical experiments indicate that
approximation errors can have the same decay rate as sparse 
approximations in wavelet bases. However, similarly to compressive sensing,
it requires that wavelet coefficients are sufficiently sparse.

Section \ref{phasecond} introduces the first main result which proves
that a non-linear rectifier acts as a filter on the phase,
and creates phase harmonics.
Section \ref{waveletphase} considers the case where
network filters are wavelets.
Section \ref{invarian} studies phase harmonic 
means and autocorrelations to capture phase dependence.
Section \ref{reconsec} shows that signals having sparse wavelet coefficients 
can be recovered from few wavelet harmonic correlations. Computations can be reproduced with a software in {\it https://github.com/kymatio/phaseharmonics}.

{\bf Notations:} 
We write $\varphi(z)$ the complex phase of $z \in \C$.
For any $x(u)$,
$\|x\|^2 = \int |x(u)|^2\,du$.
The Fourier transform of $x(u)$ is written 
$\widehat x(\om) = \int_{\R^\dd} x(u)\,e^{-i\om.u}\,du$.
The indicator function of a set $S$ is $1_S$.

\section{Phase Harmonics}
\label{phasecond}

Convolutional network architectures typically learn filters of small
support. For image and audio applications, filters in the first layer 
usually have a Fourier transform which is well
localized \citep{mallatReview}. In this case, we show that rectified
non-linearities act as a filter on the phase of the first
layer coefficients. 
Section \ref{linrepha} defines the notion of phase
by extending the real network filters into complex analytic filters. 
By applying a Fourier transform along the phase variable,
Section \ref{linsdfsa} proves that the phase filtering is a multiplicative
filter over harmonic components.
Lipschitz continuity properties of harmonic representations are
proved in Section \ref{biliplinsdfsa}.

\subsection{Rectifiers as Analytic Phase Filters}
\label{linrepha}

A one-layer convolutional network computes convolutions 
of a $\dd$-dimensional input signal $x(u)$ 
with a family of real valued filters which
we write $\{ \psi^r_m\}_m$. 
It then applies a 
pointwise non-linearity $\rho$.
The resulting first layer convolutional coefficients 
are indexed by a position $u$ and a channel $m$ 
\[
\UU x (u,m) = \rho (x \star \psi^r_m(u)). 
\]
The non-linearity is 
often chosen to be the rectifier $\rho(a) =  \max(a,0)$. 

If $x$ is positive and if $\psi^r_m$ is a low-pass
filter which averages $x$ then $\rho(x \star \psi^r_m(u)) =
x \star \psi^r_m(u)$ because $x \star \psi^r_m(u)$ remains positive.
This is not true if $\psi^r_m$ is a 
band-pass filter:
$\int \psi^r_m (u)\,du = 0$. In this case, 
we analyze the effect of the rectifier
by computing an analytic extension of $\psi^r_m$. 

\paragraph{Analytic extension}
In one dimension $u \in \R$, the analytic part $\psi_m$ 
of $\psi^r_m$ is a complex filter whose 
Fourier transform $\widehat \psi_m$ is the restriction
of $\widehat \psi^r_m$ over positive frequencies:
\begin{equation}
\label{analyt}
\widehat \psi_m (\om) = 2\, 
\widehat \psi^r_m (\om)\, 1_{[0,\infty)} (\om)~~.
\end{equation}
Since $\psi^r_m$ is real, 
$\widehat \psi^r_m (- \om) = \widehat \psi^r_m ( \om)^*$, 
so $\psi^r_m$ is the real part of $\psi_m$:
\begin{equation}
\label{real}
\psi^r_m(u) = {\rm Real}(\psi_m(u) ) .
\end{equation}
For example, if $\psi^r_m (u) = \cos(\la u + \alpha)$ with $\lambda > 0$
then its analytic part is $\psi_m (u) = e^{i (\la u + \alpha)}$.

If $u \in \R^\dd$ is multidimensional, the analytic part
is defined from the restriction
of the Fourier transform of
$\widehat \psi^r_m$ over half of the frequency space
\[
\widehat \psi_m (\om) = 2\,  \widehat \psi^r_m (\om)\, 1_{S_m} (\om)~,
\]
where $S_m$ is a half space of $\R^\dd$ whose boundary is a 
hyperplane including the frequency $0$. 
This analytic extension depends
upon the choice of $S_m$, which is not unique.
If possible, $S_m$ is chosen 
so that $\widehat \psi^r_m(\om)$ vanishes 
at the boundary of $S_m$, so that
$\widehat \psi_\la (\om)$ is not discontinuous at this boundary.
Since $\widehat \psi^r_m (-\om) = \widehat \psi^{r}_m (\om)^*$, we verify
that $\psi^r_m = {\rm Real}(\psi_m )$. 

Since $\psi^r_m = {\rm Real}(\psi_m )$,
convolutional network coefficients can be rewritten
\begin{equation}
\label{Udef}
\UU x (u,m)  = \rho( 
x \star \Real( \psi_m) (u) ) = \rho( \Real(
x \star \psi_m (u))) . 
\end{equation}

\paragraph{Phase Filter}
Let us now show that a rectifier acts as a filter on the phase of 
these complex coefficients. This is valid for any
homogeneous operator $\rho$, which means that
\[
\forall (\beta,a) \in \R^+ \times \R~,~~\rho(\beta a) = \beta\, \rho(a) .
\]
A rectifier is homogeneous, but 
an absolute value $\rho(a) = |a|$ or the identity
$\rho(a) = a$ are also homogeneous.
If $\rho$ is homogeneous then 
for any $z = |z| \,e^{i \varphi(z)} \in \C$ 
\begin{equation}
\label{homosdf}
\rho (\Real(z)) = |z|\,\hh(- \varphi(z))) ~~\mbox{with}~~
\hh(\alpha) = \rho(\cos(\alpha)).
\end{equation}
The function $\hh$ is the $2 \pi$ periodic phase filter.
For a rectifier $\hh(\alpha) = \max(\cos \alpha,0)$,
for an absolute value $\hh(\alpha) = |\cos \alpha|$, 
and for the identity $\hh(\alpha) = \cos \alpha$.

In signal processing,
the complex phase $\varphi(x \star \psi_m(u))$ is called the 
analytic phase of $x \star \psi_m^r$.
Applying (\ref{homosdf}) to $z = x \star \psi_m(u)$ proves that a
rectifier applied to output of
band-pass filters acts as a filter on the analytic phase
\begin{equation}
\label{gammaU}
\UU x(u,m)  = |x \star  \psi_m(u)|\, 
\hh(- \varphi(x \star  \psi_m(u)))~.
\end{equation}
The rectifier sets to zero all coefficients whose phases
are in $[\pi/2,3\pi/2]$.

\subsection{Phase Filtering and Selectivity}
\label{linsdfsa0}

The phase of $x \star \psi_m$ depends upon the phase
of $\psi_m$. In the first layer of a convolutional network,
there are usually several filters having
similar frequency locations but different phases. We explain the role
of the phase.

\paragraph{Phase and mean frequency}
We define the phase of real filters $\psi^r_m$ from the
complex phase of their analytic part $\psi_m$. This phase is 
evaluated at a mean frequency.
The mean frequency $\la \in \R^\dd$ of $|\widehat \psi_m|^2$ is defined by
\begin{equation}
\label{centfrew}
\la = \frac{\int \om\, |\widehat \psi_m (\om)|^2\,d\om}
{\int |\widehat \psi_m (\om)|^2\,d\om} .
\end{equation}
We suppose that $|\widehat \psi_m(\la)| \neq 0$, so 
${\widehat \psi_m (\la)} = e^{-i \alpha}\, {|\widehat \psi_m (\la)|}$. 
We introduce a filter $\psi_\la = e^{i \alpha} \psi_m$ having 
a $0$ phase at the frequency $\la$.

If $\psi_\la(u)$ is localized both
along $u$ and in frequency then $x \star \psi_\la (u)$ provides
an information about $x$ which is local both around 
the spatial variable $u$ 
and the frequency variable $\la$. These local
``space-frequency'' representations
have been studied extensively \citep{mallatbook}, 
depending upon the localization of $\psi_\la$ and $\widehat \psi_\la$.
Signal properties have been mostly analyzed
through the complex modulus $|x \star \psi_\la (u)|$. It is 
called  a spectrogram if $\psi_\la$ is 
a windowed Fourier filter and a scalogram
if $\psi_\la$ is a wavelet.

Local space-frequency representations do not easily capture the
dependencies of $x \star \psi_\la (u)$ and 
$x \star \psi_{\la'} (u)$ at different frequencies because they are
not linearly correlated. Indeed,
\[
\int \Big(x \star \psi_\la (u)\Big)\, \Big(x \star \psi_{\la'} (u)
\Big)^*\, du = (2 \pi)^{-d}
\int |\widehat x(\om)|^2\, \widehat \psi_\la (\om)\,
\widehat \psi_{\la'} (\om)^*\, d\om \approx 0
\]
if the filters are separated in
the sense that
$\widehat \psi_\la \,\widehat \psi_{\la'} \approx 0$.
This cancellation is due to phase oscillations. This is why
the phase is often removed and signal properties are usually
analyzed from the modulus $|x \star \psi_\la (u)|$. 
We shall see that
the non-linear action of a rectifier introduces correlations
while preserving phase information.

Since $\psi_\la = e^{i \alpha} \psi_m$ and $\psi^r_m = {\rm Real}(\psi_m )$, we get
\begin{equation}
\label{filter-pha}
\psi^r_m = {\rm Real}(e^{-i \alpha} \psi_\la) .
\end{equation}
We will use $\alpha$ as a free continuous
parameter in $[0,2 \pi]$ to arbitrarily modify the phase 
of $\psi^r_m$.
Let us replace the index $m$ by $(\lambda,\alpha)$ which
is more meaningful:
\begin{equation}
\label{phaseresu590sdfdsf}
\UU x(u,\la,\alpha)  = \rho\Big(\Real(e^{-i \alpha} x \star  \psi_\la(u))\Big) .
\end{equation}
Similarly to (\ref{gammaU}) we get
\begin{equation}
\label{phaseresu590sdfdsf38s}
\UU x(u,\la,\alpha) 
= |x \star  \psi_\la(u)|\, 
\hh(\alpha - \varphi(x \star  \psi_\la(u)))~,
\end{equation}
with $h(\alpha) = \rho(\cos \alpha)$. 
We study the properties of
this local space-frequency-phase representation
for arbitrary phase filters $h(\alpha)$.

Each coefficient $\UU x(u,\la,\alpha)$ 
provides information which is local along $u$ and $\la$ 
but also along the
phase $\alpha$. 
If the phase filter $\hh$ has a support $[-A,A]$ then 
$U x(u,\la,\alpha)$ in (\ref{phaseresu590sdfdsf38s})
is non-zero only at points $u$ where
$\varphi(x \star  \psi_\la(u)) \in [\alpha-A , \alpha+A]$.
For a rectifier, the support of
$\hh(\alpha) = \max(\cos \alpha,0)$ is
$[-\pi/2,\pi/2]$, which is not highly selective. 
However, this phase selection eliminates the
phase cancellation effect by ensuring that
$\UU x(u,\la,\alpha)  \geq 0$ so that the correlation
$\int \UU x(u,\la,\alpha)\,\UU x(u,\la',\alpha')\,du$
becomes non-zero even for separated frequencies
$\widehat \psi_\la\,\widehat \psi_{\la'} \approx 0$.
The properties of these correlation coefficients are 
studied in Section \ref{invarian}.
The definition of phase harmonic representations 
in (\ref{phaseresu590sdfdsf38s}) is extended
to any phase filter $h(\alpha)$ to potentially
improve selectivity along phases.

The phase $\alpha$ is similar to a local translation
parameter. In one dimension, if $\psi_\la (u) = e^{i \la u}$ then
$e^{-i \alpha} x \star  \psi_\la(u) = \widehat x(\la)\, e^{i \la (u - \alpha/\la)}$ so $\alpha/\lambda$ is a global translation parameter 
along $u$. However, if $\psi_\la (u)$ is a localized sinusoidal wave
of frequency $\la$ then 
$\alpha/\la$ induces a local translation only within the
support of $\psi_\la(u)$.
For a two-dimensional filter $\psi_\la$ 
having a localized support in $u$ and whose Fourier transform is
centered at $\la = (\la_1,\la_2)$, the phase $\alpha$ acts as
a local translation in $\R^2$ by $\alpha/|\la|$ in the direction 
of $\la$.

\paragraph{Phase selectivity}
The phase selectivity depends upon the support of $h$, which can 
be modified by a convolution of $U x(u,\la,\alpha)$ along $\alpha$.
Such a convolution is linear along $\alpha$ and identical for each
$u$ and can thus be implemented in a deep convolutional
neural networks when computing second layer coefficients.
The circular convolution of $2 \pi$ periodic functions is written
\[
a \cistar b (\alpha) = \int_{[0,2 \pi]} a(\beta)\, b(\alpha-\beta) \, 
d \beta .
\]
Computing the convolution of 
$\UU x(u,\la,\alpha)$ in (\ref{phaseresu590sdfdsf38s})
by $g(\alpha)$ along $\alpha$ gives
\begin{equation}
\label{confI}
U x(u,\la,.)\cistar g (\alpha) = 
 |x \star \psi_\la(u)|\, \hh \cistar g ( \alpha - \varphi(x \star \psi_{\la} (u)) .
\end{equation}
It changes the phase filter $h$ into $h \cistar g$.
The following theorem proves that $\hh \cistar g$ 
may have an arbitrarily
narrow support and thus
become highly selective in phase, for an appropriate
choice of $g$.

\begin{theorem}
\label{filtering}
If $\hh(\alpha) = \max(\cos \alpha,0)$ or 
$\hh(\alpha) = |\cos \alpha|$ then for
any $\epsilon > 0$ there exists 
a  bounded filter $g$ such that $\hh \cistar g$ is differentiable,
$\pi$ periodic, supported in $[-\epsilon,\epsilon]$ modulo $\pi$,
with $\int_0^{2\pi} \hh \cistar g (\alpha)\,d\alpha = 2$, and
\begin{equation}
\label{gammaU00}
\lim_{\epsilon \rightarrow 0} \hh \cistar g (\alpha) = \delta (\alpha\, {\rm mod}\, \pi). 
\end{equation}
\end{theorem}

{\it Proof:} 
We show that we can define $g$ so that  
$\hh \cistar g$ is a $\pi$ periodic cubic box 
spline supported in 
$[-\epsilon,\epsilon]$ with 
$\widehat {\hh \cistar g} (0) = \int_0^{2 \pi} \hh \cistar g (\alpha)\, d\alpha = 2$. 
One can verify that
the Fourier transform of this box spline is
\begin{equation}
\label{consdf}
\widehat {\hh \cistar g} (k) = \left\{
\begin{array}{ll}
2 & \mbox{if $k = 0$}\\
{2 \sin^4(k \epsilon / 4)}\, {(k \epsilon/4)^{-4}} & \mbox{if $k \neq 0$ is even}\\
0 & \mbox{if $k$ is odd} .
\end{array}
\right.
\end{equation}

For a rectifier and an absolute value, 
$\widehat \hh(k)$ is non-zero when $k$ is even. 
Since $\widehat {h \cistar g} = \widehat \hh \, \widehat g$
we obtain (\ref{consdf})
by defining $\widehat g(k) = 0$ if $k$ is odd and
$\widehat g(k) = \widehat \hh_g (k) / \widehat \hh (k)$ if 
$k$ is even. 
We prove that $g$ is bounded by verifying that 
$\sum_k |\widehat g(k)| < \infty$. The analytical expressions of 
$\widehat h(k)$ for a rectifier and an absolute value 
are given by (\ref{nsdfousdfs}) and (\ref{nsdfousdfs2}). They imply
that $|\widehat g(k)| = O(|k|^{-2})$ and hence that $g$ is bounded.

The restrictions of $\hh \cistar g$ to $[-\pi/2,\pi/2]$ and
$[\pi/2,3 \pi/2]$ are bounded functions of integrals equal to $1$.
Their supports are respectively $[-\epsilon,\epsilon]$ and
$[\pi-\epsilon,\pi+\epsilon]$. When $\epsilon$ goes to $0$ the restrictions
of $\hh \cistar g$ to $[-\pi/2,\pi/2]$ and
$[\pi/2,3 \pi/2]$ thus converge to
$\delta(\alpha)$ and $\delta(\alpha-\pi)$,
which proves (\ref{gammaU00}).
$\Box$

This theorem proves that for a rectifier or an absolute value,
one can improve arbitrarily the phase 
selectivity through convolutions along phases. 
The theorem proof gives a possible but non-unique
choice for the filters $g$.
For wavelet filters and images,
Section \ref{waveletphase} shows that lines of constant phase
$\alpha$ define the geometry of multiscale edges.

\subsection{Phase Harmonics and Frequency Transpositions}
\label{linsdfsa}

This section proves that the
phase filtering becomes a multiplication
over a representation of phase
harmonics. These phase harmonics correspond to
non-linear frequency transpositions.

\paragraph{Phase harmonics}
We separate the linear part of $\UU$ from the non-linear 
phase filtering by factorizing
\begin{equation}
\label{gamma4}
\UU  = H\, W .
\end{equation}
The linear operator $W$ computes convolutions with 
analytic filters of mean frequency $\la$ having a
zero phase at $\la$, where $\la$ belongs to a set $\Lambda$:
\[
\forall \la \in \Lambda~~,~~W x(u,\la) = x \star \psi_\la(u) ~,
\]
The phase filter $H$ is defined by 
\begin{equation}
\label{gamsdfnsdf}
\forall z \in \C~~,~~H z(\alpha) = 
|z|\, \hh(\alpha-\varphi(z)) .
\end{equation}
Since $\alpha$ is a phase translation over $[0,2\pi]$, we prove that 
a Fourier transform relatively to $\alpha$
transforms $H$ into a multiplication over 
a sequence of phase harmonics defined below.

\begin{definition}
The phase harmonics of a complex number $z \in \C$ is  
a sequence defined for all $k \in \Z$ by
\[
[z]^k = |z|\, e^{i k \varphi(z)}~.
\]
\end{definition}

As opposed to $z^k$ which applies the exponent $k$ to the modulus and
to the phase, the phase harmonic $[z]^k$ transforms the phase only. 
As a result Lemma \ref{Lipschhar} will prove 
that $[z]^k$ is Lipschitz for all $k \in \Z$ 
whereas $z^k$ is not Lipschitz for 
$|k| \neq 1$.  The Fourier coefficients 
of a $2 \pi$ periodic function $\hh(\alpha)$ are written
\[
\widehat \hh (k) = \frac 1 {2 \pi} \int_0^{2 \pi}
\hh(\alpha) \, e^{-i k \alpha }\, d \alpha~.
\]
The following proposition proves that 
the Fourier transform of
$H z(\alpha)$ is proportional to the phase harmonics of 
$z$. It results 
that the Fourier transform of $\UU x(u,\la,\alpha)$ relatively
to $\alpha$ is proportional to the phase harmonics
of $x \star \psi_\la(u)$.

\begin{proposition}
\label{Fousdfnsdf}
For all $z \in \C$, the Fourier transform of $H z(\alpha)$ is
\begin{equation}
\label{phaseresu5}
\widehat H z(k) = \widehat \hh(k)\, [z]^{-k} ~.
\end{equation}
The Fourier transform of $\UU x(u,\la,\alpha)$ along $\alpha$ is
\begin{equation}
\label{phaseresu590sdf}
\widehat \UU x (u,\la,k) = \widehat \hh(k)\, [x \star \psi_\la(u)]^{-k}~.
\end{equation}
\end{proposition}

Equation (\ref{phaseresu5}) and (\ref{phaseresu590sdf})
are directly obtained by computing the
Fourier transform of (\ref{gamsdfnsdf}) and (\ref{phaseresu590sdfdsf38s})
along $\alpha$. The Fourier transform
$\widehat \UU x (u,\la,k)$ is a local space-frequency-harmonic representation
where $k$ is the harmonic exponent. The phase filter produces
a harmonic multiplier $\widehat \hh(k)$, which typically attenuates high order
harmonics. 

If $\UU x(u,\la,\alpha)$ is calculated
with a homogeneous non-linearity $\rho$ then $h(\alpha) = \rho(\cos \alpha)$. 
If $\rho(a) = a$ then
$\hh(\alpha) = \cos \alpha$ so $\widehat \hh(k) = 1/2$ if $k = \pm 1$ and 
$\widehat \hh (k) = 0$ otherwise. This is usually not an appropriate choice
because it eliminates all high-order harmonics.
If $\rho(a) = \max(a,0)$ then $\hh(\alpha) =  \max(\cos \alpha,0)$ and a direct calculation
of Fourier integrals gives
\begin{equation}
\label{nsdfousdfs}
\widehat {\hh}(k) = \left\{
\begin{array}{ll}
\frac{-(i)^k} {\pi (k-1) (k+1)} & \mbox{if $k$ is even}\\
\frac{1} 4 & \mbox{if $k= \pm 1$}\\
0 & \mbox{if $|k|>1$ is odd}
\end{array}
\right. .
\end{equation}
One can verify that
any real homogeneous operator can be written
$\rho(a) = \gamma\, \max(a,0) + \beta\, a$
so $\widehat \hh(k) = 0$ when $|k| > 1$ is odd. 
We may however define other phase filters for which $\widehat h(k) \neq 0$
for $|k|>1$ odd.
If $\rho(a) = |a|$ then $\hh(\alpha) =  |\cos \alpha|$ and hence
\begin{equation}
\label{nsdfousdfs2}
\widehat {\hh}(k) = \left\{
\begin{array}{ll}
\frac{-2(i)^k} {\pi (k-1) (k+1)} & \mbox{if $k$ is even}\\
0 & \mbox{if $k$ is odd}
\end{array}
\right. .
\end{equation}
Because the absolute value is even and thus 
loses sign information, we have $\widehat \hh(1) = \widehat \hh(-1) = 0$.

\paragraph{Frequency transposition}
We show that phase harmonics produce frequency
transpositions. They multiply the frequencies of
$x \star \psi_\la$ by a factor $k$ and hence perform 
a non-linear dilation by $k$ of the Fourier 
transform of $x \star \psi_\la$,  
without affecting its spatial localization. 

The derivative of the phase $\varphi(x \star \psi_\la (u))$
of $x \star \psi_\la (u)$ at $u$ is called the analytic instantaneous frequency
in signal processing \citep{mallatbook}. 
It corresponds to the perceived frequency of a sound at a given time.
Since $[x \star \psi_\la(u) ]^k = |x \star \psi_\la(u)| e^{i k \varphi(x \star \psi_\la (u))}$, it has a phase derivative which is $k$ times
larger. However, its modulus is unchanged so $x \star \psi_\la$ and
$[x \star \psi_\la ]^k$ have the same localization along $u$.
For time signals, 
a transformation which modifies the instantaneous frequency
but does not modify the time distribution is called a frequency
transposition, by analogy to music transpositions.
For example, shifting a musical score by one octave multiplies by two the ``frequencies'' of all  musical notes, but it does not
change the tempo and the melody. 

A frequency transposition performs a non-linear
dilation in the Fourier domain. 
The Fourier support of $x \star \psi_\la$ is included in the 
support of $\widehat \psi_\la$, which is centered at $\la$. 
In a first approximation,
the Fourier transform of $[x \star \psi_\la]^k$ is non negligible over
a domain centered in $k \la$ whose width is approximately
$k$ times the support width of $\widehat \psi_\la$. 
This is proved by the following proposition, 
under very restrictive hypotheses. 
A ball in $\R^\dd$ is written $B(c,r) = \{u \in \R^\dd~:~|u - c| \leq r \}$.

\begin{proposition}
If $|x \star \psi_\la (u)|$ and $e^{i \varphi(x \star \psi_\la (u))}$ 
have a Fourier transform respectively supported in
$B(0,\Delta)$ and $B(\lambda,\Delta)$ then 
$[x \star \psi_\la]^k$ has a Fourier transform supported in
$B(k\lambda , (|k|+1) \Delta)$. 
\end{proposition}

{\it Proof:} Since $[x \star \psi_\la]^k = |x \star \psi_\la|\, 
( e^{i  \varphi(x \star \psi_\la (u))})^k$, its Fourier transform is the
convolution of the Fourier transform of $|x \star \psi_\la|$ and
$k$ successive convolutions of the Fourier transform of 
$e^{i  \varphi(x \star \psi_\la (u))}$. 
The Fourier support of $[x \star \psi_\la]^k$
can be derived from the Fourier supports of 
$|x \star \psi_\la|$ and $e^{i  \varphi(x \star \psi_\la (u))}$
because the convolutions of two functions
included in $B(c,r)$ and in $B(c',r')$ is a function included in
$B(c+c',r+r')$. $\Box$. 

If $\widehat \psi_\la$ is supported in
$B(\la, \Delta)$ then $x \star \psi_\la$ has a Fourier 
transform supported in $B(\la, \Delta)$ but in general 
$|x \star \psi_\la|$ and $e^{i  \varphi(x \star \psi_\la (u))}$ do not have
a Fourier transform of compact support. 
Indeed, $|x \star \psi_\la|$ may be singular if 
$x \star \psi_\la(u)$ has a zero-crossing, which produces
a Fourier transform with a slow asymptotic decay.
The hypotheses of this proposition are thus very restrictive.
For wavelet filters, Section \ref{waveletphase} verifies numerically that
the Fourier transform of $[x \star \psi_\la]^k$ is essentially
dilated by $k$ for a white noise signal.

\subsection{Bi-Lipschitz Continuity and Inversion}
\label{biliplinsdfsa}

This section considers general
phase filters $\hh(\alpha)$ and gives
bi-Lipschitz bounds on $H$ depending upon $\hh$.
We derive that $\UU$ is also bi-Lipschitz
and invertible for appropriate filters $\psi_\la$.

For any $2 \pi$ periodic $g(\alpha)$, we have
\[
\| g \|^2 = \frac 1 {2 \pi} \int_0^{2 \pi} |g(\alpha)|^2\, d \alpha = \sum_{k \in \Z} |\widehat g(k)|^2. 
\]
We first prove that each phase harmonic $[z]^k$ is
Lipschitz, even if $|k| \neq 1$,
as opposed to usual integer power exponents $z^k$.

\begin{lemma}
\label{Lipschhar}
For any $(z,z') \in \C^2$ and any $k \in \Z$
\begin{equation}
\label{nsdoifsasd}
|[z]^ k - [z']^k| \leq \max(1,|k|)\, |z - z'|~.
\end{equation}
\end{lemma}

{\it Proof:}
The inequality (\ref{nsdoifsasd}) is verified if $k = 0$ because
\[
|[z]^k - [z']^k| = ||z| - |z'|| \leq |z-z'|.
\]
For $k \neq 0$ and $\epsilon = \varphi(z') - \varphi(z)$, let us define
\begin{eqnarray*}
f(|z|,|z'|,\epsilon) & = & \frac{|[z]^ k - [z']^k|^2} {|z - z'|^2} = \frac{| |z| - |z'| e^{i k \epsilon}|^2} {| |z| - |z'| e^{i \epsilon}|^2}\\
& = & \frac{| |z|^2 + |z'|^2 - 2 |z| |z'| \cos(k \epsilon)|} 
{| |z|^2 + |z'|^2 - 2 |z| |z'| \cos( \epsilon)|} 
\end{eqnarray*}
One can verify that
\[
\sup_{|z|,|z'|,\epsilon} f(|z|,|z'|,\epsilon) = |k|^2 ,
\]
which implies (\ref{nsdoifsasd}). This is done
by proving that for $|z|$ and $|z'|$ fixed, the maximum of $f$ is 
reached when $\epsilon$ tends to $0$, and when $\epsilon$ tends to zero
$f(|z|,|z'|,\epsilon)$ is maximized when $|z| = |z'|$, with a supremum 
equal to $|k|^2$.  $\Box$

The following theorem applies this lemma to 
prove that $H$ is bi-Lipschitz.

\begin{theorem}
\label{biolsdfsda}
For any $(z,z') \in \C^2$
\begin{equation}
\label{nsdfsdfa}
\| H z \| = \|\hh\|\, |z|~,
\end{equation}
and
\begin{equation}
\label{nsdfsdfa8sdfs}
\sqrt 2\, |\widehat \hh(1)|\,|z - z'| \leq
\| H z - H z' \| \leq \kappa\,|z - z'| 
\end{equation}
with 
\begin{equation}
\label{nsdfsdfa8sdfs8sd}
\kappa^2 = |\widehat \hh(0)|^2 + \sum_{k\in \Z} k^2\, |\widehat \hh(k)|^2 = 
\frac 1 {4 \pi^2} \Big(\int_0^{2 \pi} \hh(\alpha)\, d \alpha \Big)^2+ \frac 1 {2 \pi}
\int_0^{2 \pi} |\hh'(\alpha)|^2\, d \alpha  .
\end{equation}
For a rectifier, 
$\|\hh \| = 1/2$, $\widehat \hh(1) = 1/4$ and $\kappa = \sqrt{1/4 + 1/\pi^2}$. 
\end{theorem}

{\it Proof:} Since $\widehat H z(k) = \widehat \hh(k)\, [z]^{-k}$,
we first prove (\ref{nsdfsdfa}) by observing that
\[
\|H z\|^2 = \sum_{k=-\infty}^{+\infty} |\widehat \hh(k)|^2 |[z]^{-k}|^2 =
|z|^2 \sum_{k=-\infty}^{+\infty} |\widehat \hh(k)|^2 = |z|^2\, \|\hh\|^2.
\]
To verify (\ref{nsdfsdfa8sdfs}) let us compute
\begin{equation}
\label{gam9nsdf09s}
\| H z - H z' \|^2 = \sum_{k=-\infty}^{+\infty} 
|\widehat \hh(k)|^2\, |[z]^{-k} - [z']^{-k}|^2 .
\end{equation}
Since $\hh(\alpha)$ is real, 
$|\widehat \hh(k)| = |\widehat \hh(-k)|$. 
Restricting the sum to $k = \pm 1$ gives the lower bound
\[
2 \, |\widehat \hh(1)|^2\, |z - z'|^2 \leq \| H z - H z' \|^2 .
\]

We obtain the upper bound inequality of (\ref{nsdfsdfa8sdfs}) 
inserting the Lipschitz inequality
(\ref{nsdoifsasd}) of Lemma \ref{Lipschhar}
in (\ref{gam9nsdf09s}) with 
\begin{equation}
\label{nsdfsdfa8sdfs8sd0}
\kappa^2 = 
|\widehat \hh(0)|^2 + \sum_{k\in \Z} k^2\, |\widehat \hh(k)|^2 ,
\end{equation}
Since $\widehat \hh(0) = 1/{2 \pi} \int_0^{2 \pi} \hh(\alpha)\, d\alpha$
and the Fourier transform of $\hh'(\alpha)$ is $i k \widehat \hh(k)$,
applying the Plancherel formula on (\ref{nsdfsdfa8sdfs8sd0}) proves
(\ref{nsdfsdfa8sdfs8sd}). 
A direct calculation gives the values of $\|\hh\|$ and $\kappa$
for a linear rectifier where $\hh(\alpha) = \max(\cos \alpha,0)$.
$\Box$

In the particular case where the phase filter can be written
$\hh(\alpha) = \rho(\cos \alpha)$ then
the Lipschitz constant $\kappa$ in 
(\ref{nsdfsdfa8sdfs8sd}) is 
finite if $\rho$ has bounded derivatives.
For a rectifier the lower and upper Lipschitz constants of the
theorem are
respectively about $0.35$ and $0.69$ and thus within a factor $2$.
In this case, one can prove that 
the upper bound Lipschitz constant is not tight and can
be reduced to $1/2$ \citep{ZhangMallat}.

The following proposition reviews a standard
condition which guarantees 
that $W$ is invertible and stable, by imposing
that the Fourier
domain is uniformly covered by the filters $\widehat \psi_\la$.

\begin{proposition}
If there exists $0 \leq \eta < 1$ such that for almost all $\om \in \R^\dd$
\begin{equation}
\label{littndoisdf}
(1 - \eta)^2 \leq \frac 1 2 
\sum_{\laL} \Big( |\widehat \psi_\la (\om)|^2 + |\widehat \psi_\la (-\om)|^2 \Big) \leq (1 + \eta)^2 ~
\end{equation}
then for all $x \in \Ld(\R^\dd)^2$ real
\begin{equation}
\label{nsdfsdf89sdf5at}
x = \sum_{\laL} \Real( x \star \psi_\la \star \overline \psi_{\la}) ,
\end{equation}
where the support of $\widehat {\overline  \psi_\la}$ is the same as the
support of $\widehat \psi_\la$ and
\[
\widehat {\overline \psi_\la}(\om) = 
\frac{\widehat { \psi_\la}(\om)^*} 
{\sum_{\laL}  |\widehat \psi_\la (\om)|^2} .
\] 
Moreover, for all $(x,x') \in \Ld(\R^\dd)^2$ real
\begin{equation}
\label{nsdfsdf89sdf}
(1 - \eta)^2 \,\|x - x'\|^2 \leq \|W x - W x'\|^2 \leq (1 + \eta)^2 \|x - x'\|^2~.
\end{equation}
\end{proposition}

{\it Proof:} 
Since $x$ is real, $\widehat x(-\om) = \widehat x(\om)^*$. 
Equation (\ref{nsdfsdf89sdf5at}) is proved by taking the Fourier transform on both side and observing that 
\[
\frac 1 2 \sum_{\laL} \Big( \widehat \psi_\la (\om)\, \widehat {\overline \psi_\la} (\om) +
\widehat \psi_\la (-\om)^*\, \widehat {\overline \psi_\la} (-\om)^*\Big) = 1 . 
\]
Since $W$ is linear, (\ref{nsdfsdf89sdf}) can be
proved by setting $x' = 0$. Observe that
\[
\|W x\|^2 = \sum_{\la \in \Lambda} \|x \star \psi_\la \|^2 = 
\frac 1 {(2 \pi)^\dd} \, \sum_{\la \in \Lambda} \int |\widehat x(\om)|^2\, |\widehat \psi_\la (\om)|^2\, d\om .
\]
Inverting the
sum and the integral and 
inserting (\ref{littndoisdf}) in this equation proves (\ref{nsdfsdf89sdf}) for $x' = 0$. $\Box$

The following corollary derives that 
$\UU = H W$ is invertible and bi-Lipschitz. We write
\[
\|\UU x \|^2 = \frac 1 {2 \pi} \sum_{\la \in \Lambda} \int_0^{2 \pi}
\int_{\R^\dd} |\UU x(u,\la,\alpha)|^2 \, du  \,d \alpha~.
\]

\begin{corollary}
\label{recnsdf}
If condition (\ref{littndoisdf}) is satisfied 
for $0 \leq \eta < 1$ and $\widehat h(1) \neq 0$ 
then for all $x \in \Ld (\R^\dd)$ real
\begin{equation}
\label{nsdfsdf89sdf5at76s}
x(u) = 
\sum_{\laL} \Real\Big( \frac{1}{\widehat h(1)^*}\,
\widehat \UU x(.,\la,1)^* \star \overline \psi_{\la}(u) 
\Big) ,
\end{equation}
\begin{equation}
\label{nsdfsdfanusdf}
\|h\|\, (1 - \eta) \,\|x\| \leq \| \UU x \| \leq \|h\|\, (1 + \eta)\, \|x\|~
\end{equation}
and for all $(x,x') \in \Ld(\R^\dd)^2$ real
\begin{equation}
\label{nsdfsdfanusdfu9sdf}
\sqrt 2 \, |\widehat h(1)|\, (1 - \eta)\,\|x - x' \| \leq
\| \UU x -  \UU x'\| \leq \kappa\, (1 + \eta)\, \|x - x' \|~.
\end{equation}
\end{corollary}

{\it Proof:} It results from (\ref{phaseresu590sdf}) that
$\widehat \UU x(u,\la,1)= \widehat h(1) \,x \star \psi_\la (u)^*$, together with(\ref{nsdfsdf89sdf5at}) it implies (\ref{nsdfsdf89sdf5at76s}).

Since $\UU = H \, W$, (\ref{nsdfsdf89sdf}) together with (\ref{nsdfsdfa}) proves (\ref{nsdfsdfanusdf}), and 
(\ref{nsdfsdf89sdf}) together with (\ref{nsdfsdfa8sdfs}) proves
(\ref{nsdfsdfanusdfu9sdf}).
$\Box$

\section{Wavelet Transforms}
\label{waveletphase}

Wavelets are dilated filters $\psi_\la$
which separate signal variations at multiple scales. 
They provide sparse representations of large
classes of signals and images. 
Sections \ref{1D} and \ref{2D} review the 
properties of complex analytic wavelet transforms and introduce a new
bump wavelet for numerical calculations. 

\subsection{Analytic Wavelets for 1D Signals}
\label{1D}

A one-dimensional wavelet transform is computed by convolutions with
dilated wavelets.
Analytic wavelets $\psi$ have a
Fourier transform $\widehat \psi(\om)$ which
is zero at negative frequencies.
This section introduces a new analytic bump wavelet, 
which is used in numerical calculations. It provides sparse
representations of piecewise regular signals. 

We impose that $\widehat \psi$ is real valued which implies 
that the real part of $\psi$ is even and its imaginary part is odd. 
Let $\xi$ be the mean frequency of $\psi$ 
according to (\ref{centfrew}). We also suppose that 
$\widehat \psi(\xi) > 0$.
A wavelet transform with $Q$ scales per octave 
is calculated by dilating
$\psi$ by $2^{j/Q}$, where $j$ is an integer:
\[
\psi_{\la} (u) = 2^{-j/Q} \, \psi (2^{-j/Q}\, u)~~\mbox{and hence}~~
\widehat \psi_{\la} (\om) =  \widehat \psi (2^{j/Q}\, \om)~.
\]
The mean frequency of $\psi_\la$ is
\[
\la = 2^{-j/Q} \xi .
\]
If the energy of $\widehat \psi(\om)$ is concentrated in an interval
centered at $\xi$ of radius $\beta \xi$ then
$\widehat \psi_\la$ is concentrated in an interval
centered at $\la$ of radius $\beta \la$.

A real filter of phase $\alpha$ is defined by 
${\rm Real}(e^{-i \alpha} \psi_\la)$. 
The phase $\alpha$ is thus a symmetry 
parameter which makes the transition from even to odd filters,
and which changes the filters sign when adding $\pi$.

Let $2^J$ be the maximum scale. 
Scales larger than $2^J$ are carried by
a low-pass filter $\psi_0$ centered
at the frequency $\la = 0$ and dilated by $2^J$. 
We use a Gaussian filter whose Fourier transform is
\begin{equation}
\label{gaussian}
\widehat \psi_0 (\om) =  \exp \Big (  - \frac {|\om|^2} {2 \sigma_J^2} \Big) .
\end{equation}
\begin{figure}
\begin{center}
 \includegraphics[width=1.5in]{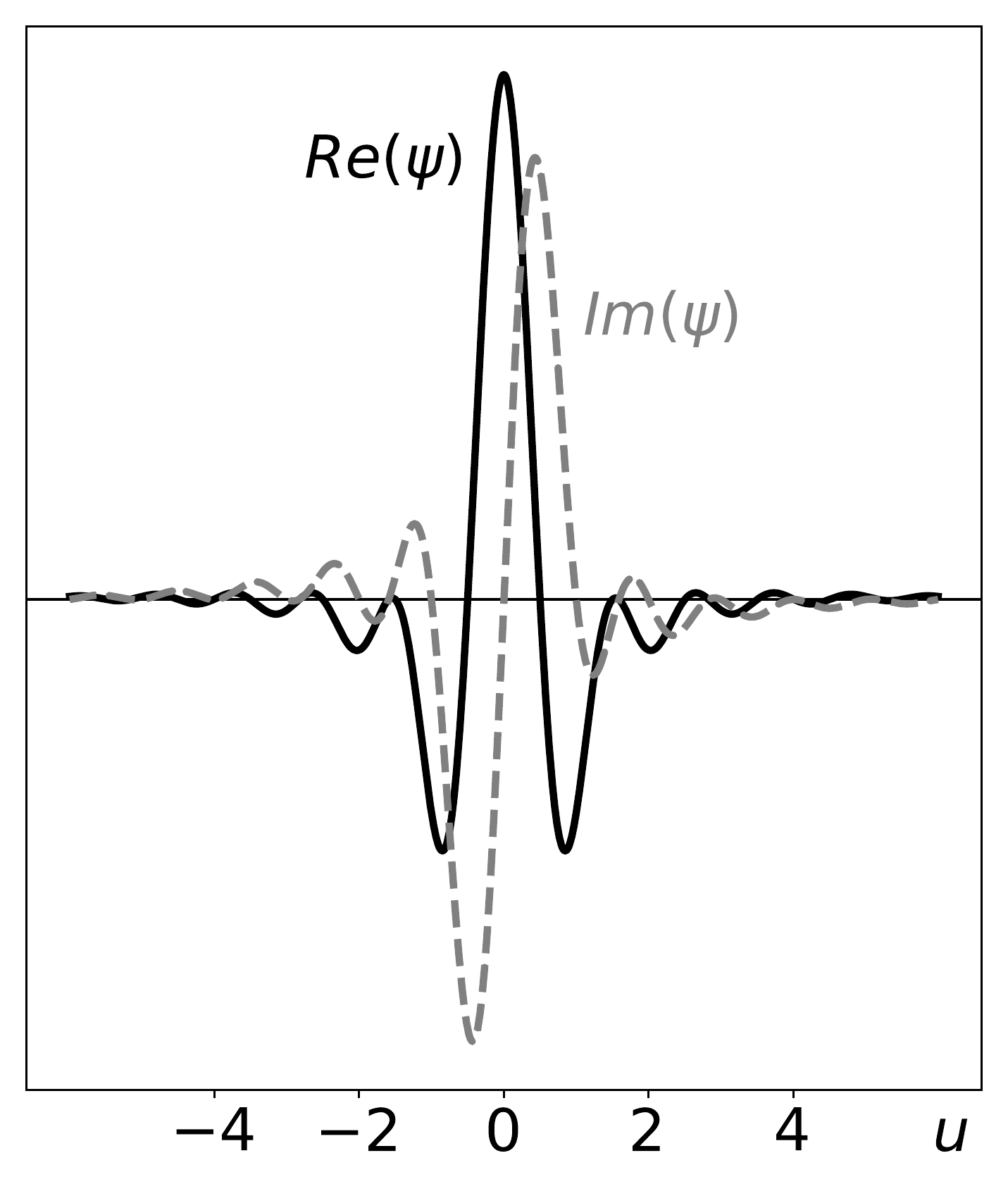}
  \includegraphics[width=1.5in]{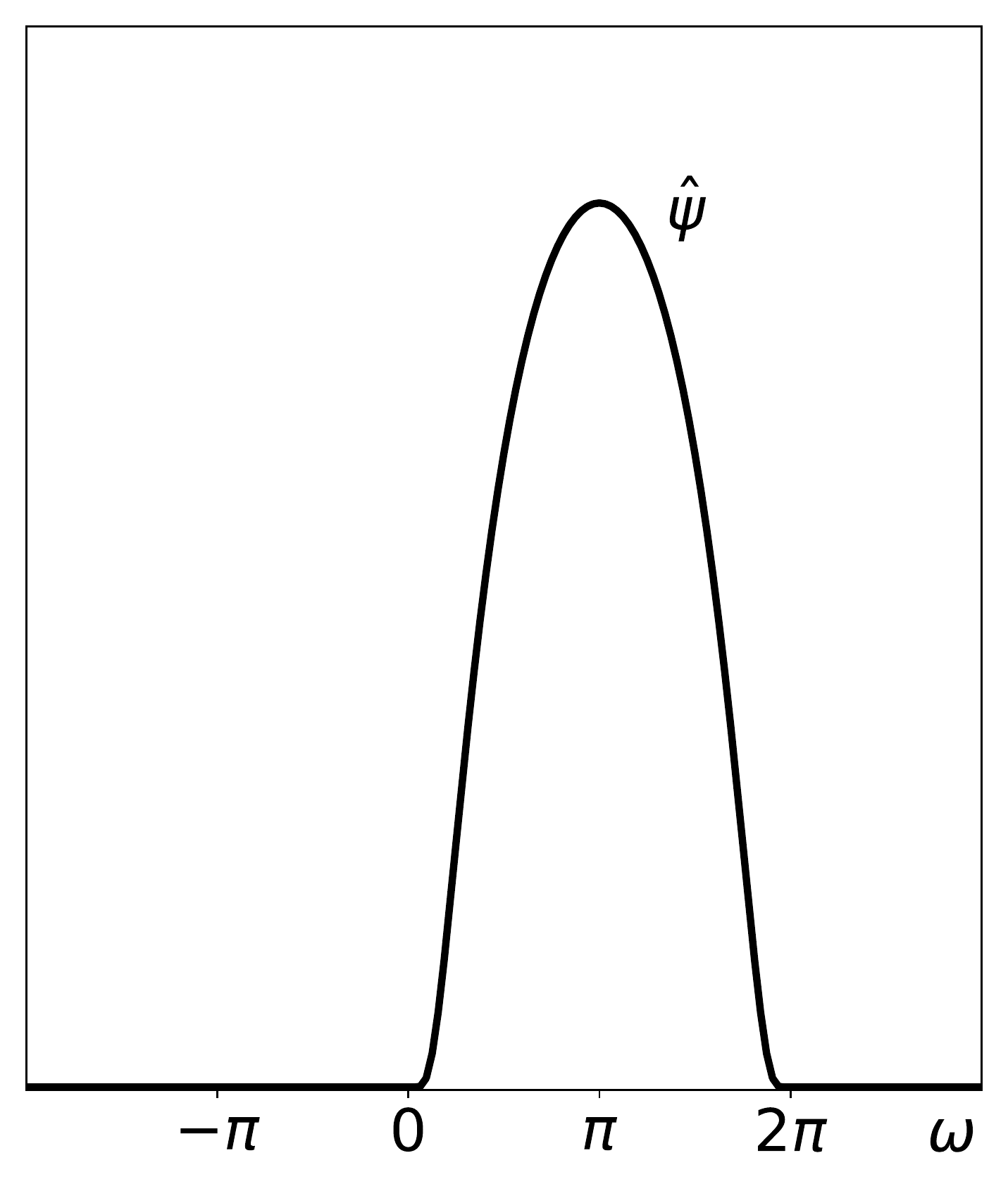}\\
(a)~~~~~~~~~~~~~~~~~~~~~~~(b)
\end{center}
\caption{(a): The real and imaginary parts
of a bump wavelet $\psi(u)$, for $Q = 1$, are shown 
with a full and a dashed line.
(b): Fourier transform $\widehat \psi (\om)$. }
\label{bumpwave}
\end{figure}

\begin{figure}
\begin{center}
        \includegraphics[width=2in]{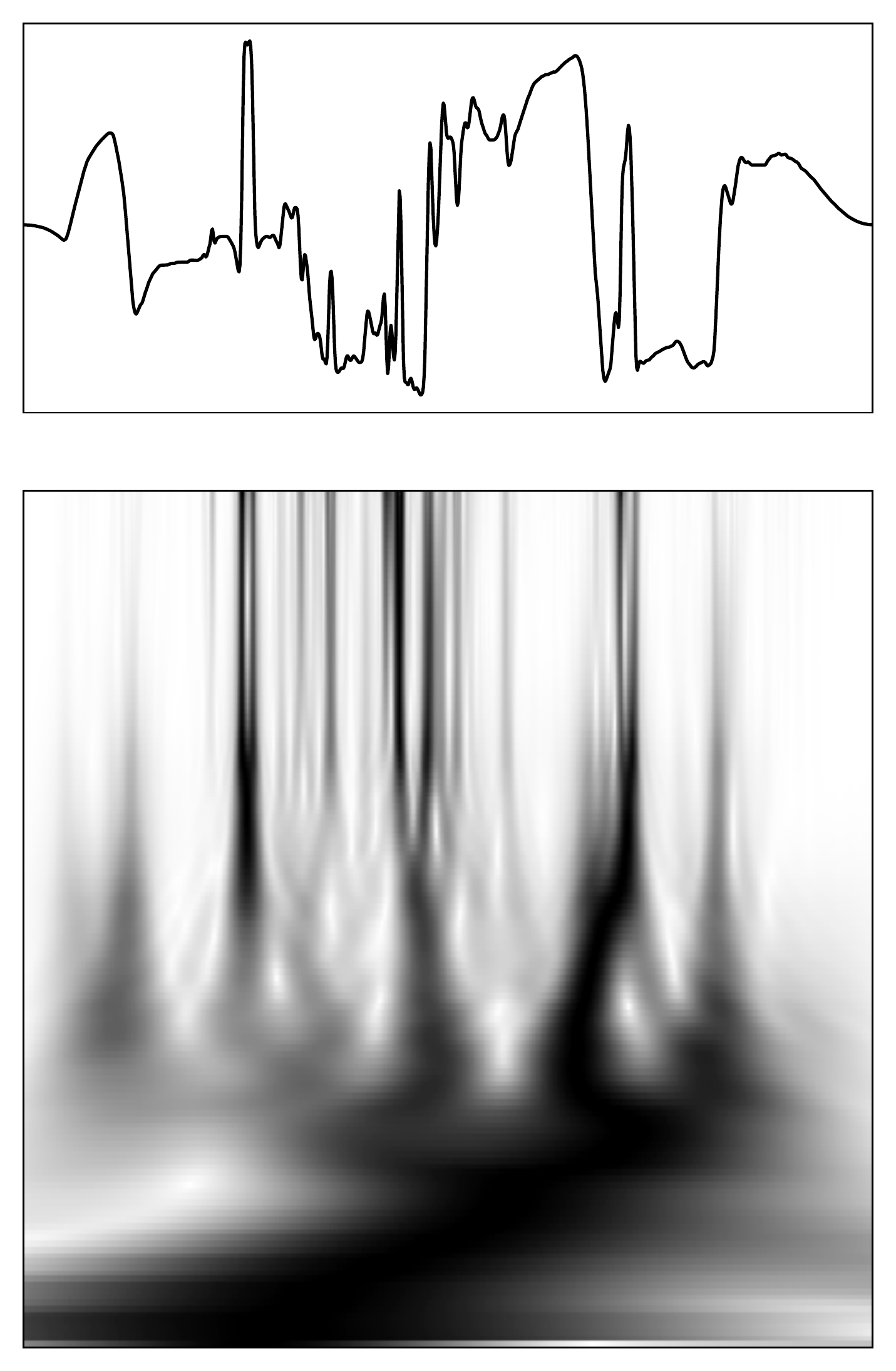}
        \includegraphics[width=2in]{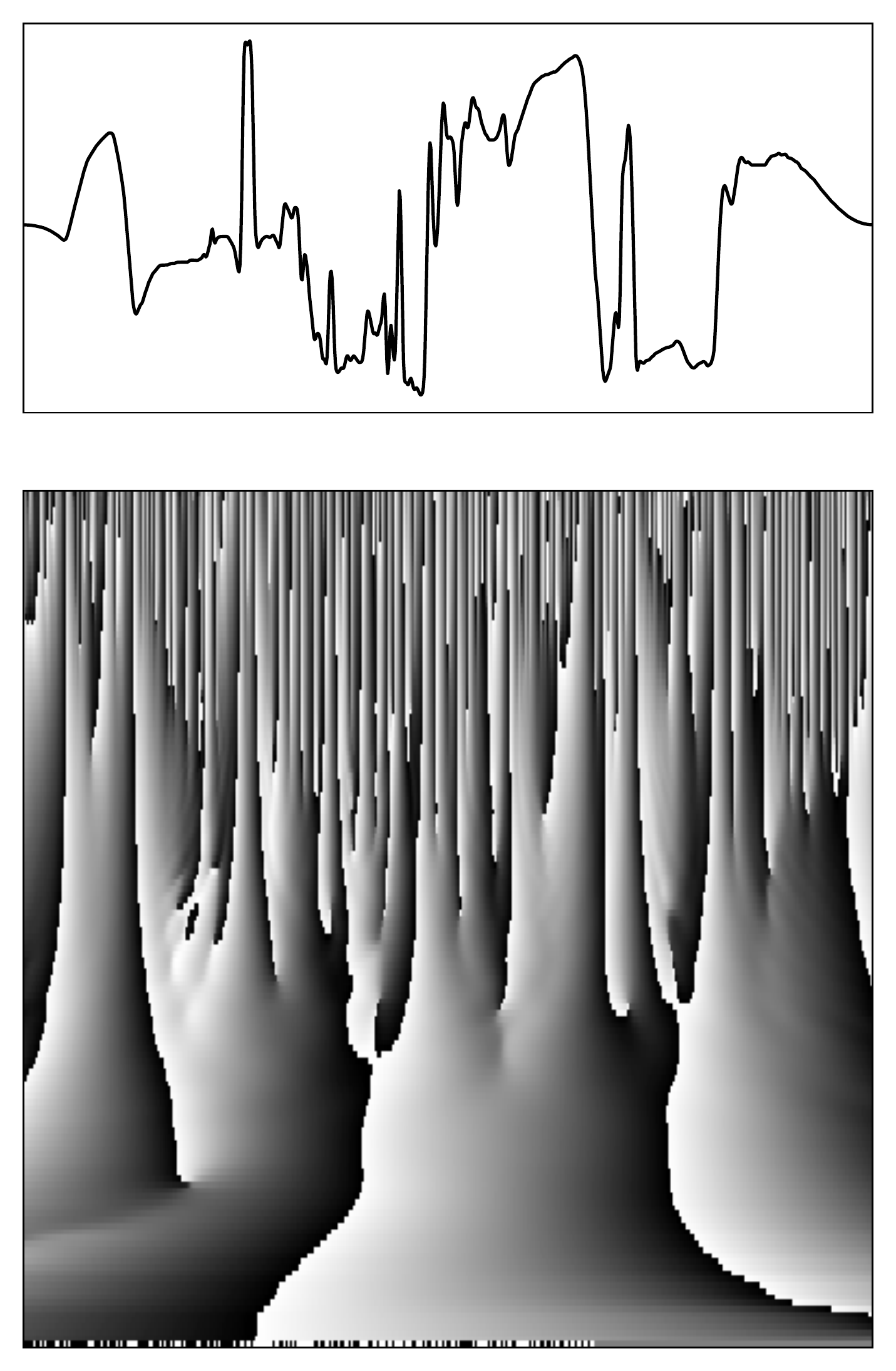}\\
(a)~~~~~~~~~~~~~~~~~~~~~~~~~~~~~~~~~~(b)
\end{center}
\caption{Top: original signal $x(u)$ as a function of $u$. 
(a): Wavelet transform modulus $|x * \psi_{\la}(u)|$ 
with $Q = 16$ scales per octave, as a function of $(u,\log_2 \la)$ 
along the horizontal and vertical axes.
White and black points correspond respectively
to small and large amplitudes. 
(b): Complex phase $\varphi (x * \psi_{\la}(u))$ as a function of 
$(u,\log_2 \la)$.}
\label{sig1D} 
\end{figure}

We consider an analytic wavelet 
whose Fourier transform $\widehat \psi(\om)$ is
a regular window centered at a frequency $\xi > 0$
\begin{equation}
\label{bump0}
\widehat \psi(\om) = c\, g\Big(\frac{\om - \xi}{\xi}\Big),
\end{equation}
where $g(\om)$ has a support in $[-1,1]$ with $g(1) = g(-1) = 0$, so that 
$\widehat \psi(\om) = 0$ for $\om \leq 0$.
The window $g$ is chosen so that $\psi$ is well localized both in
the spatial and Fourier domains, which has a tendency to produce 
wavelet coefficients which are more sparse.
A bump window is an infinitely differentiable 
approximation of a Gaussian having
a support equal to $[-1,1]$ \citep{Bump}: 
\begin{equation}
\label{bump}
g(\om ) =  \exp \Big (  \frac {-|\omega|^2} {1 - |\om|^2} \Big)\, 1_{(-1,1)} (\om)~.
\end{equation}
It defines
a compact support $\widehat \psi(\om)$ which is ${\bf C}^{\infty}$.
The resulting {\it bump} wavelet $\psi$ is
a Schwartz class ${\bf C}^{\infty}$ analytic function, with 
a decay faster than any rational function.
Since all derivatives of $\widehat \psi$ vanish at $\om = 0$, $\psi$
has an infinite number of vanishing moments:
\[
\forall k \in \N~~,~~\int u^k\, \psi(u)\, du = 0~.
\]
Vanishing moments are important so that wavelet coefficients 
$x \star \psi_\la (u)$ are small in domains of $u$ where $x$ is regular
\citep{mallatbook}. 

For wavelets, the Fourier transform condition (\ref{littndoisdf}) is called the
Littlewood-Paley inequality. It guarantees that
the wavelet transform is invertible and stable, with bounds which
depend upon $\eta$. To minimize $\eta$,
the constants in (\ref{gaussian}) and 
(\ref{bump0}) are chosen to be
\[
\sigma_J = 2^{\frac{-0.550}{Q}}  \, 2^{-J+1}\,\xi~~\mbox{and}~~
c  = (1.34 \sqrt{Q} - 0.05)^{-1}.
\]
In numerical applications, we choose $\xi = 0.85\,\pi$.
For these bump wavelets,
$\eta = 0.091$ when $Q=1$ and $\eta \leq 0.035$ when $Q \geq 2$.
Figure \ref{bumpwave} shows the real and imaginary parts of $\psi$ 
for $Q = 1$, as well as its Fourier transform.

Figure \ref{sig1D} gives the modulus and the
phase of the wavelet transform of a one-dimensional signal, calculated with
a bump wavelet, with $Q = 16$ scales per octave.
Fine scales correspond to high frequencies
$\la$. Large modulus coefficients $|x \star \psi_\la (u)|$ are
sparse. 
They are located in the neighborhood of sharp signal transitions.
The phase $\varphi(x \star \psi_\la (u))$ 
gives a local symmetry information on the transition of $x \star \psi_\la$
at $u$. Since the real and imaginary parts of $\psi_\la$ are respectively
symmetric and antisymmetric, 
the variations of 
$x \star \psi_\la$ is locally symmetric in the neighborhood
of $u$ if $\varphi(x \star \psi_\la (u)) = 0$ and 
antisymmetric if this phase is $\pi/2$. Changing the
sign of $x \star \psi_\la$
adds $\pi$ to the phase. When the phase is $\pi/2$ modulo $\pi$ then
the real part of $x \star \psi_\la (u)$ has a zero-crossing.
Grossmann et. al.  
It was shown in \citep{Grossmann} that
lines of constant phase across scales capture properties of instantaneous
frequencies. 
Zero-crossings of real wavelet transforms have been studied to reconstruct
signals and detect the position of sharp transitions \citep{ZhangMallat}.

\begin{figure}
\center
\includegraphics[width=4in]{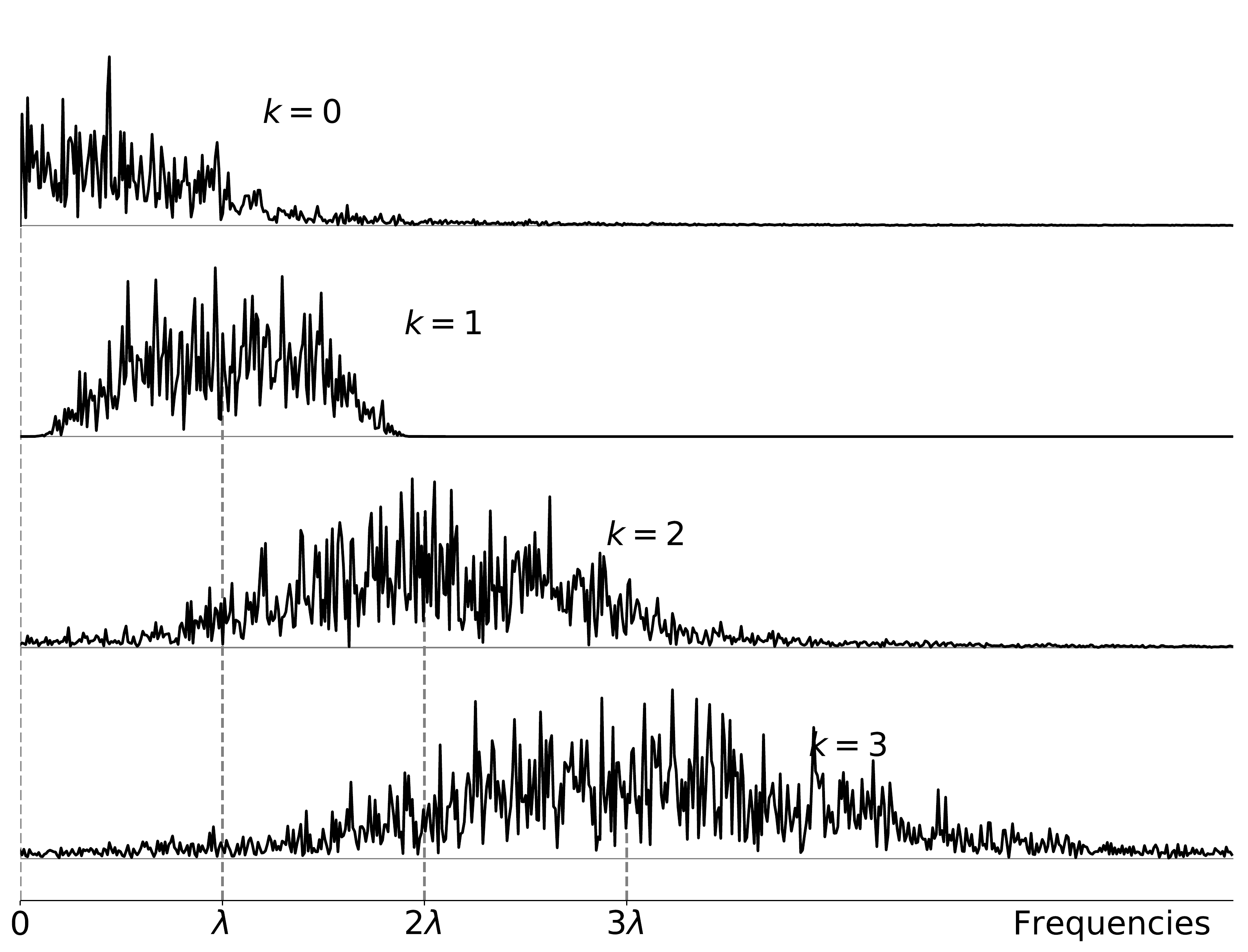}
\caption{Modulus of the Fourier transform of $[x \star \psi_\la]^k$,
for a signal $x$ which is a Gaussian white noise and a bump wavelet
$\psi_\la$, as a function of  frequencies.
The center frequency $\la$ is fixed and $k$ increases from top to bottom. The Fourier supports are approximately dilated by $k$.}
\label{Fourierradius}
\end{figure}

Section \ref{linsdfsa} explains that harmonics
$[x \star \psi_\la]^k = |x \star \psi_\la|\, e^{i k \varphi(x \star \psi_\la)}$
perform a frequency transposition which approximately dilates the
Fourier transform $x \star \psi_\la$ by a factor $k$.
This is illustrated by Figure \ref{Fourierradius} for 
a Gaussian white noise $x$ filtered by bump wavelet $\psi_\la$.
For $k = 0,1,2,3$, the  modulus of the Fourier transform of 
$[x \star \psi_\la]^k$ has an energy 
centered at $k \la$ and concentrated over a domain 
dilated by $\max(k,1)$.

\subsection{Complex Steerable Wavelets for Images}
\label{2D}

In two dimensions, we define wavelets by dilating and rotating a
complex analytic wavelet $\psi(u)$ for $u \in \R^2$. 
Its Fourier transform $\widehat \psi(\om)$ has a 
support included in the right half plane of $\R^2$. 
This wavelet is rotated in $\R^2$
by $r_{\theta}$ along several angles
$\theta$, and dilated at dyadic scales $2^j$:
\[
\psi_{\la} (u) = 2^{-2j} \psi (2^{-j} r_{-\theta} u) ~~\mbox{and hence}~~
\widehat \psi_\la (\om) = \widehat \psi(2^j r_\theta \om)~.
\]
We use $L$ angles $\theta = \pi \ell / L$ for $-L/2 < \ell \leq L/2$.
If $\xi$ is the center frequency of $\psi$ then the center frequency
of $\psi_\la$ is
\[
\la = 2^{-j} r_{-\theta}\, \xi~.
\]

Wavelets are computed up to a maximum scale $2^J$.
Scales larger than $2^J$ are carried by
a low-pass filter $\psi_0$ centered
at the frequency $\la = 0$ and dilated by $2^J$. As in one dimension,
we use a Gaussian filter whose Fourier transform is
\begin{equation}
\label{gaussian2}
\widehat \psi_0(\om) = \exp \Big (  - \frac {| \om|^2} {2 \sigma_J^2} \Big)
~.
\end{equation}

\begin{figure}
\begin{center}
        \includegraphics[width=5in]{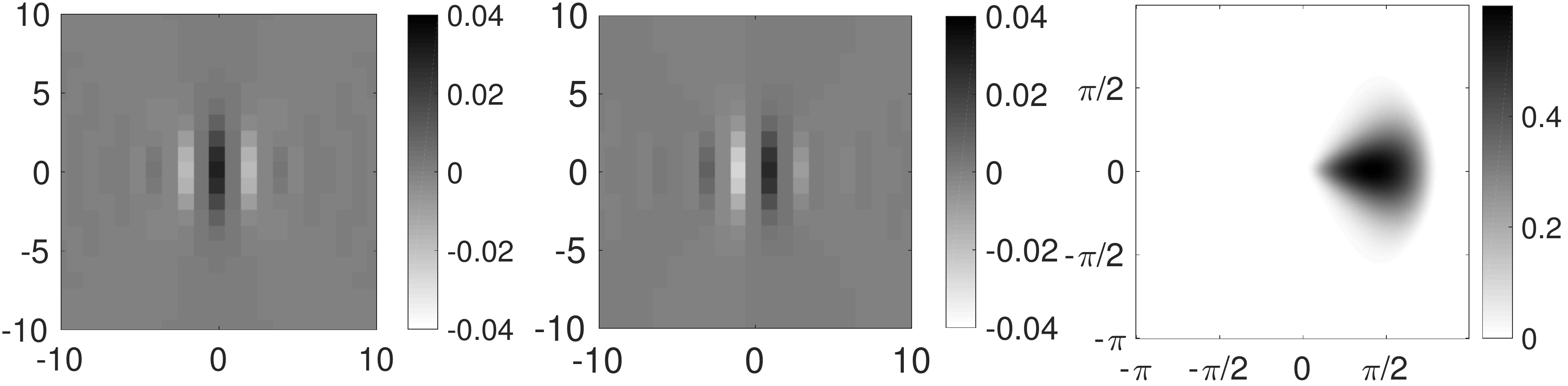}\\
(a)~~~~~~~~~~~~~~~~~~~~~~~~~~~~(b)~~~~~~~~~~~~~~~~~~~~~~~~~~~~(c)
\end{center}
\caption{(a): Real part of the two-dimensional bump wavelet $\psi(u)$ with 
$L=8$. 
(b): Imaginary part of $\psi(u)$. (c): Fourier transform $\widehat \psi(\om)$.}
\label{bumpwave2}
\end{figure}

\begin{figure}
\begin{center}
	\includegraphics[width=1.2in]{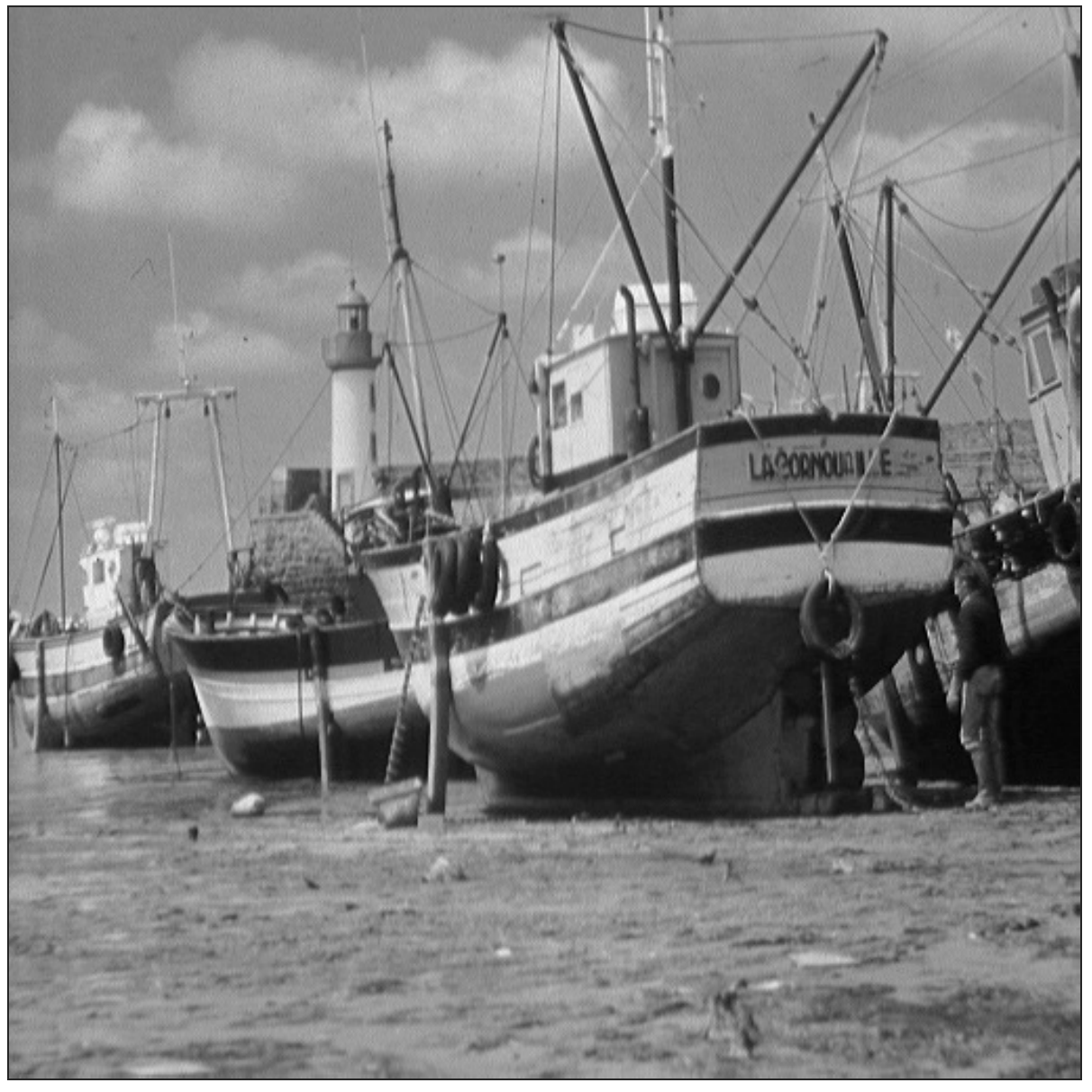}\\
        \includegraphics[width=1.2in]{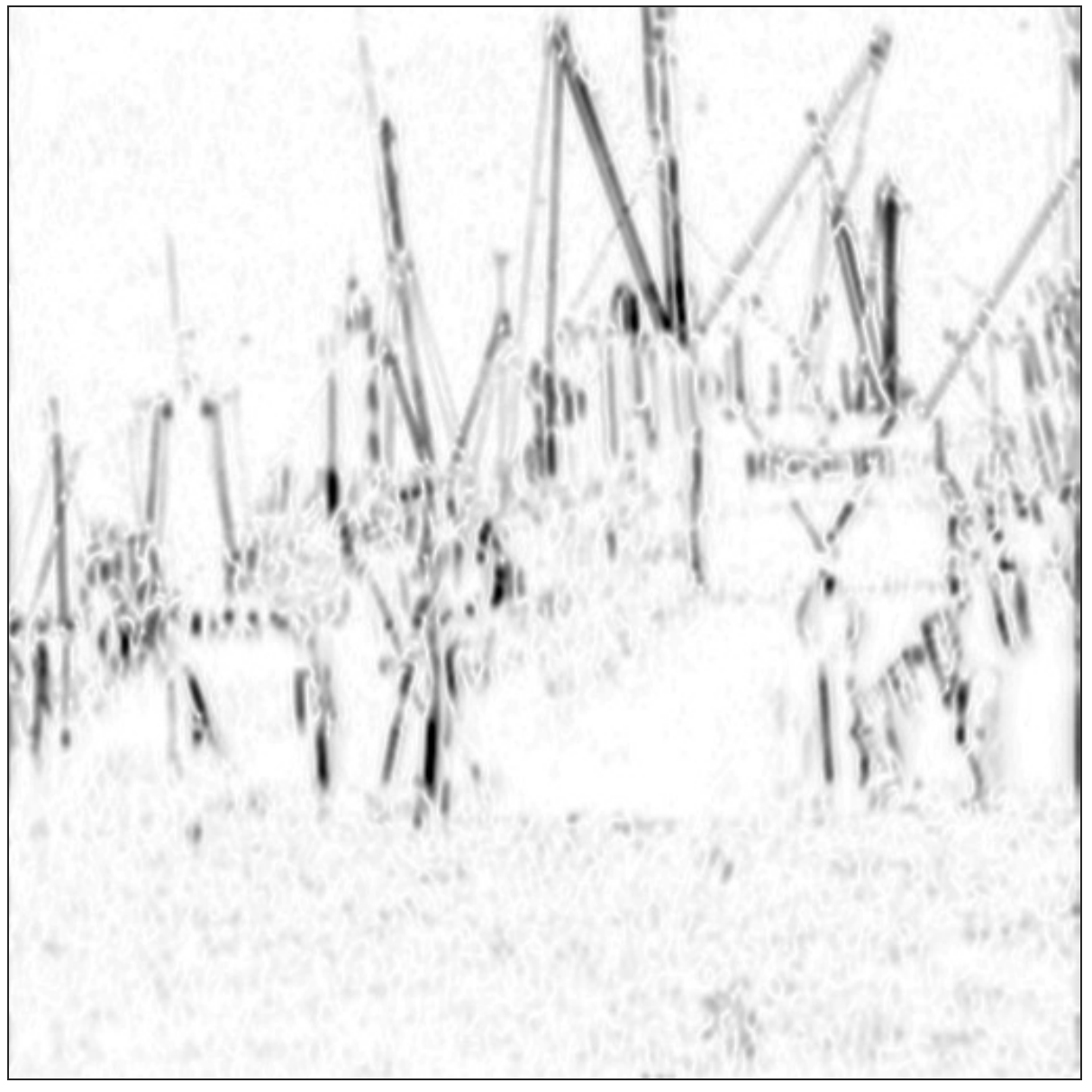}
        \includegraphics[width=1.2in]{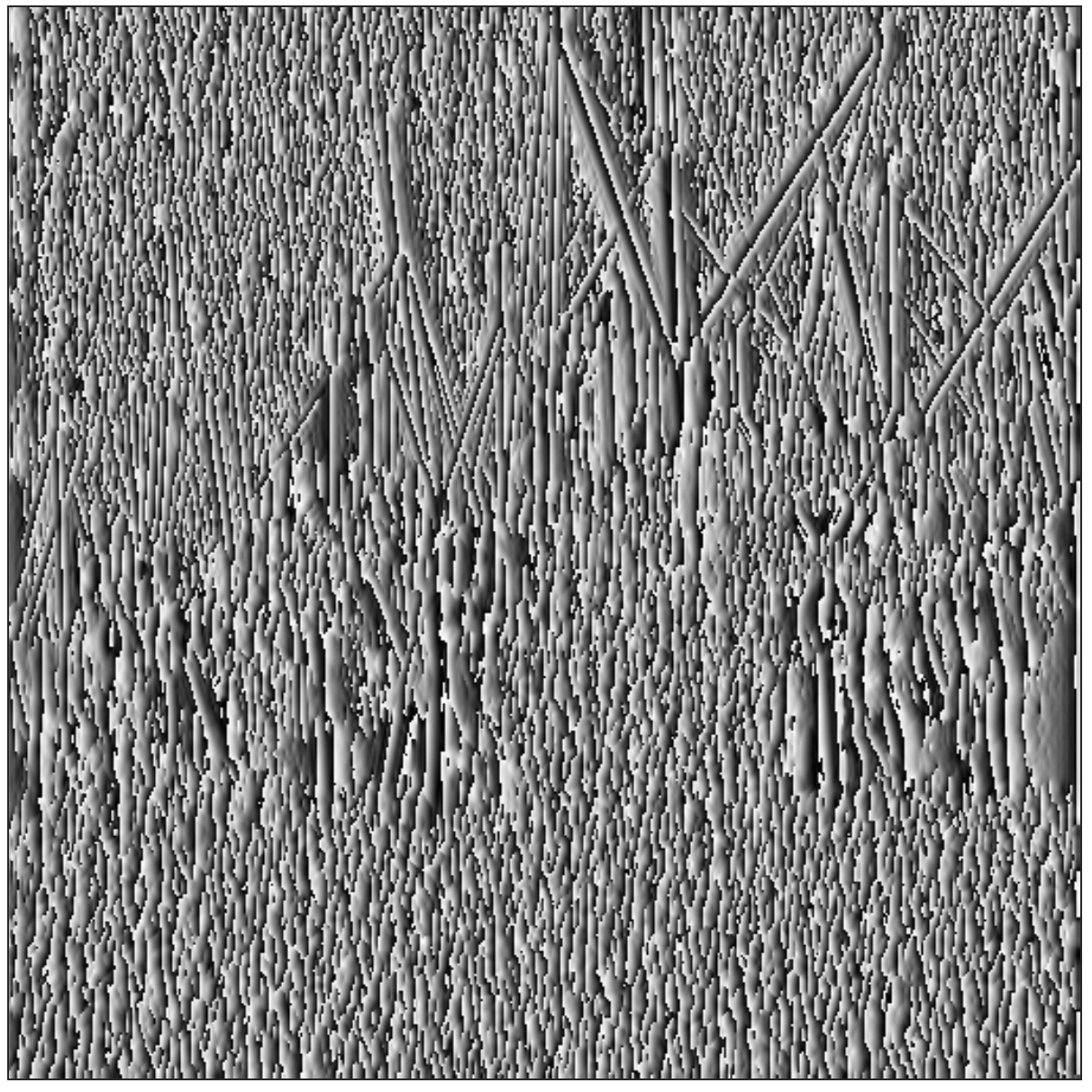}
        \includegraphics[width=1.2in]{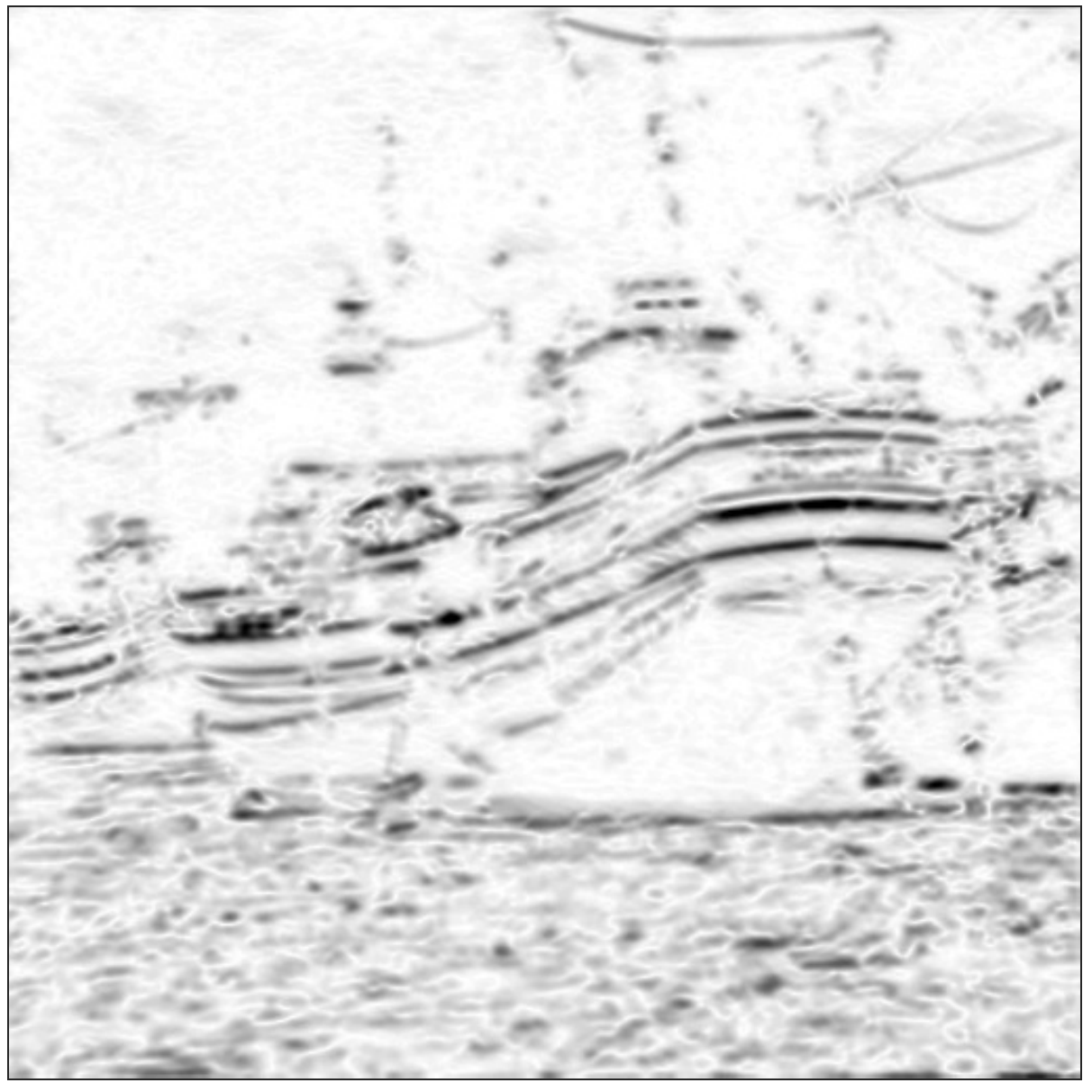}
        \includegraphics[width=1.2in]{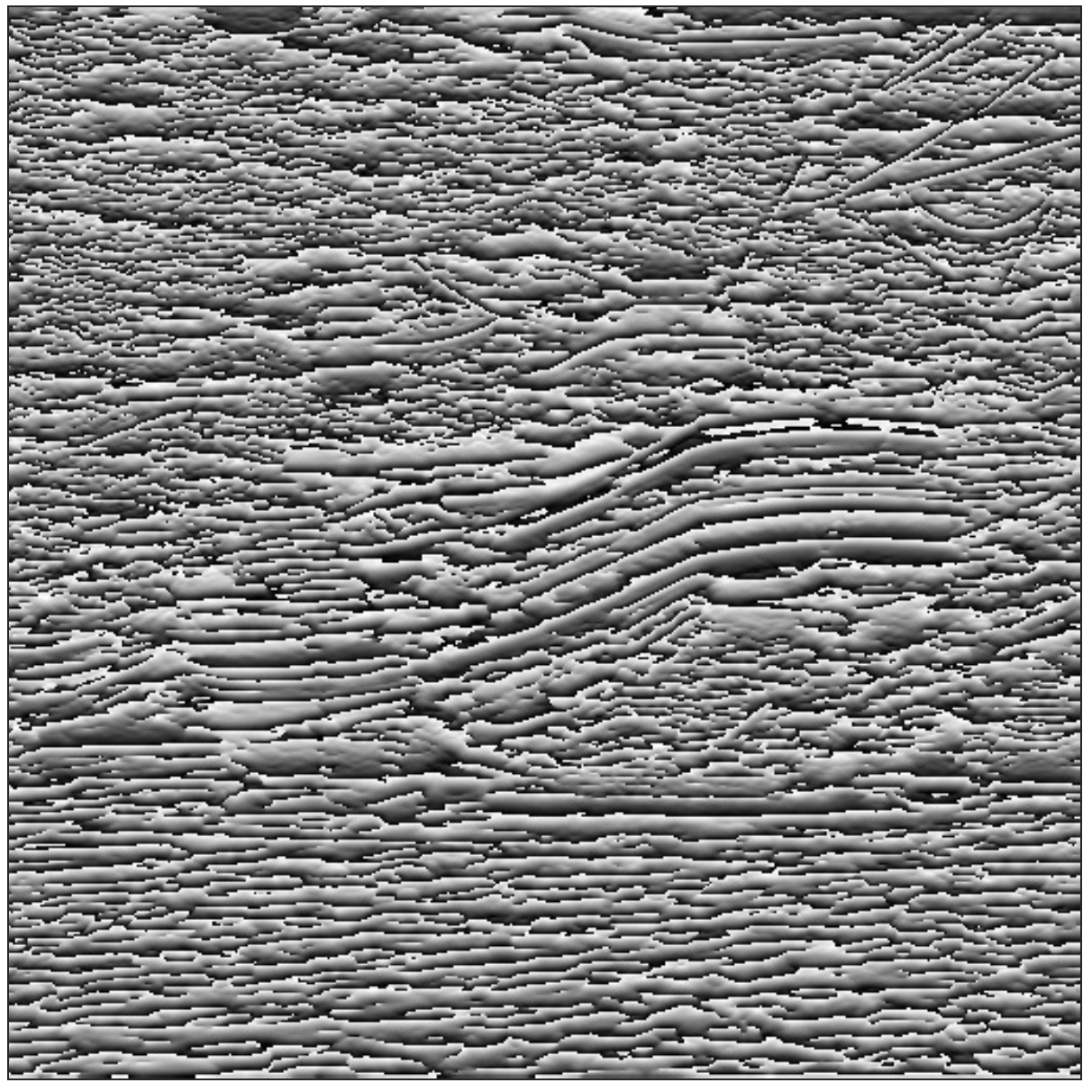}\\
        \includegraphics[width=1.2in]{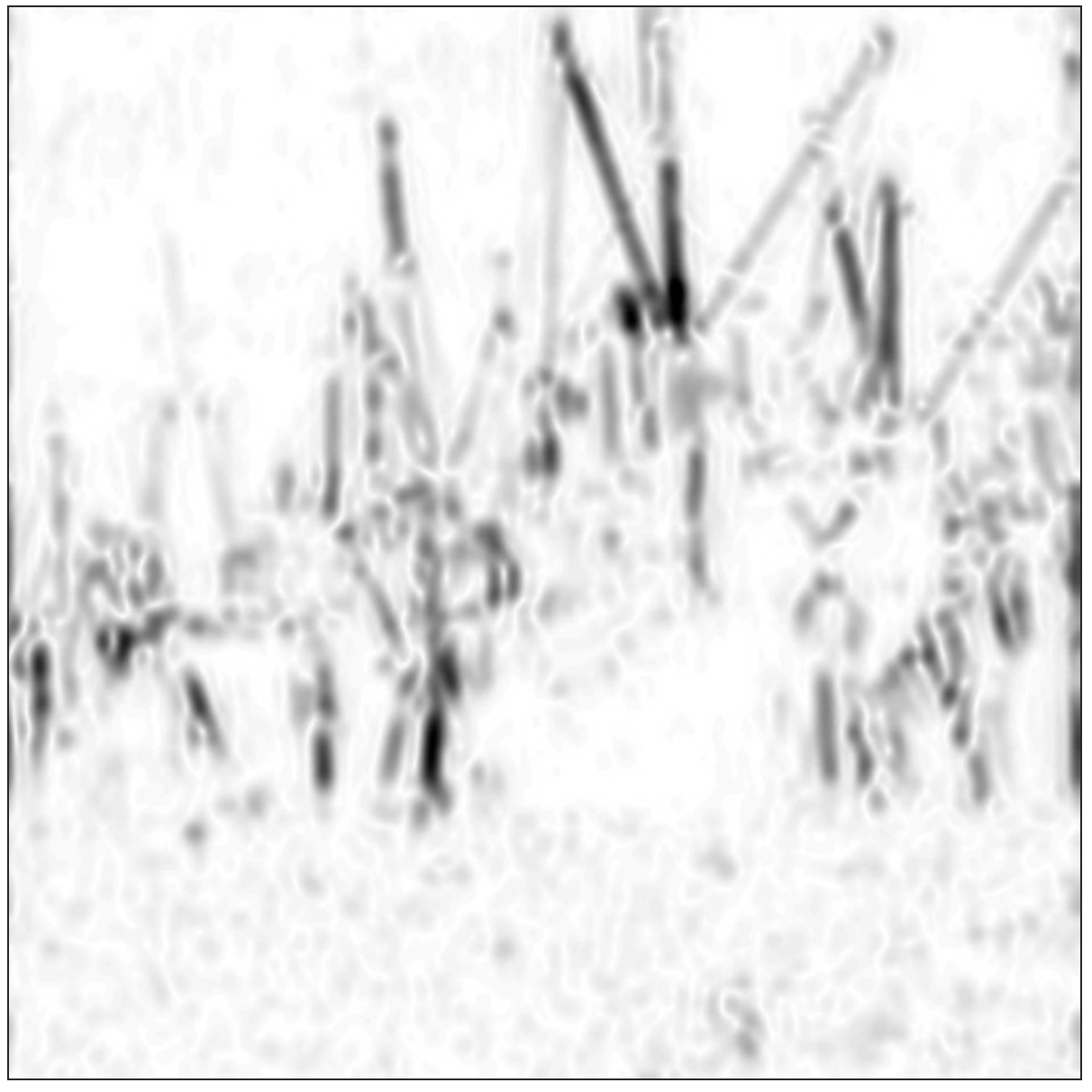}
        \includegraphics[width=1.2in]{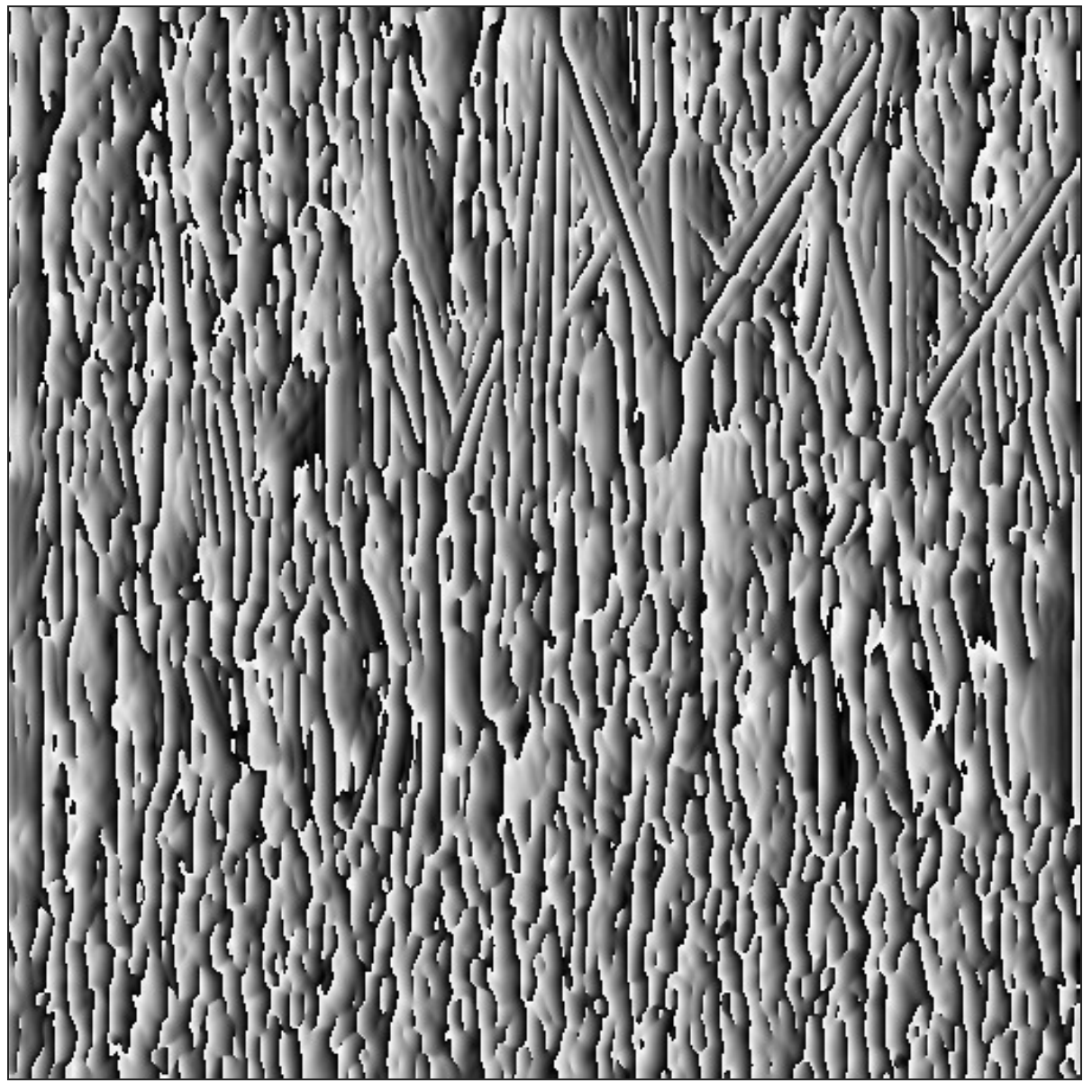}
        \includegraphics[width=1.2in]{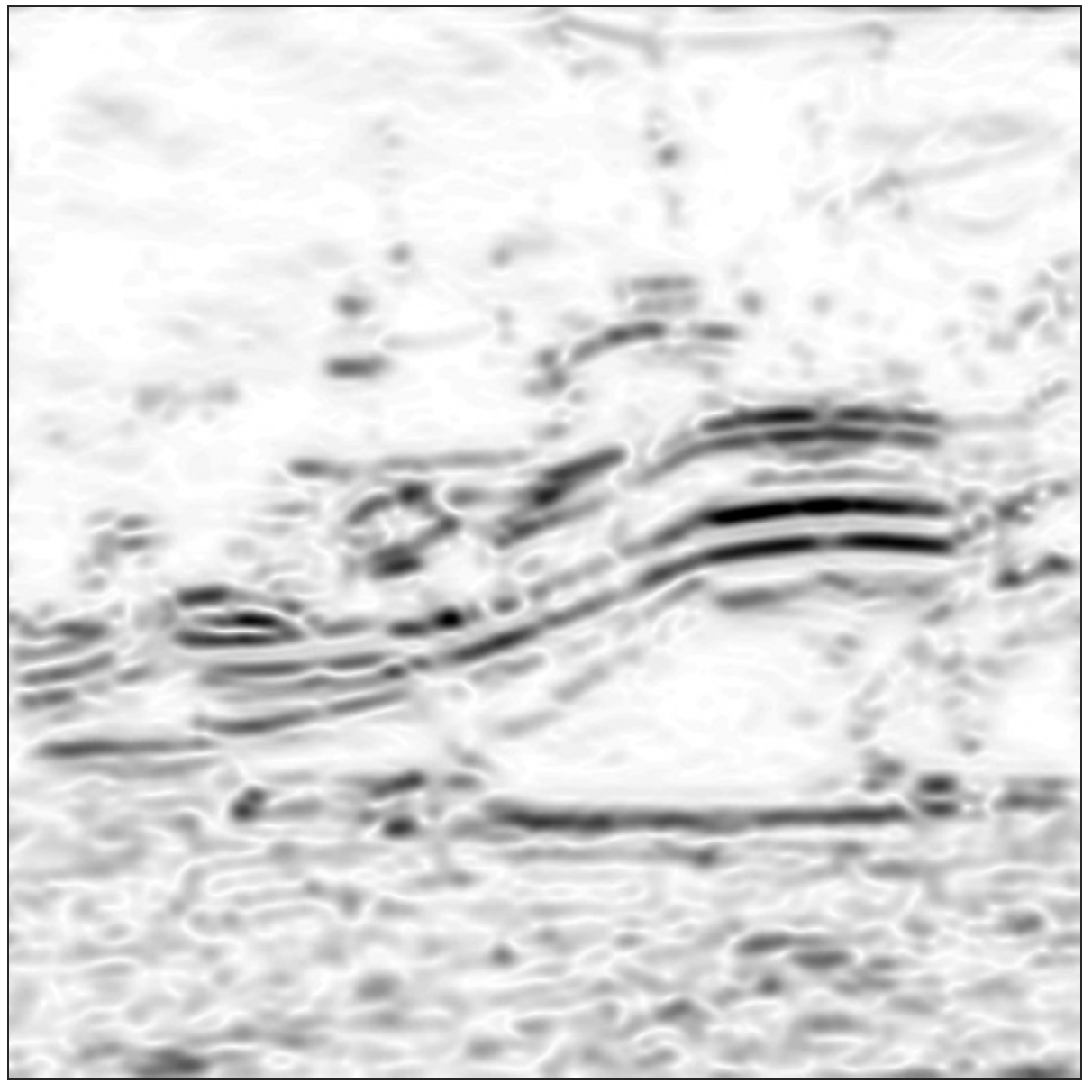}
        \includegraphics[width=1.2in]{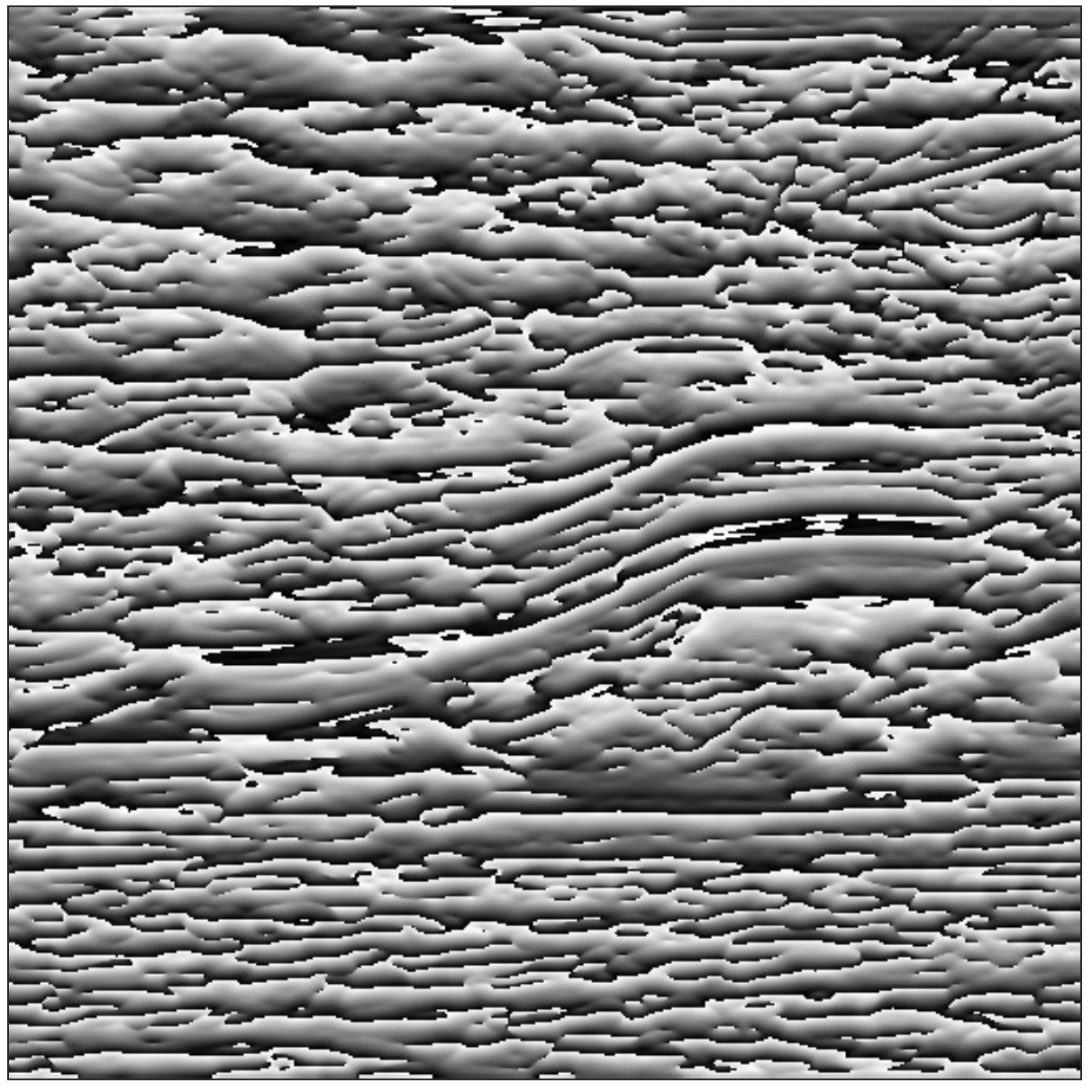}\\
        \includegraphics[width=1.2in]{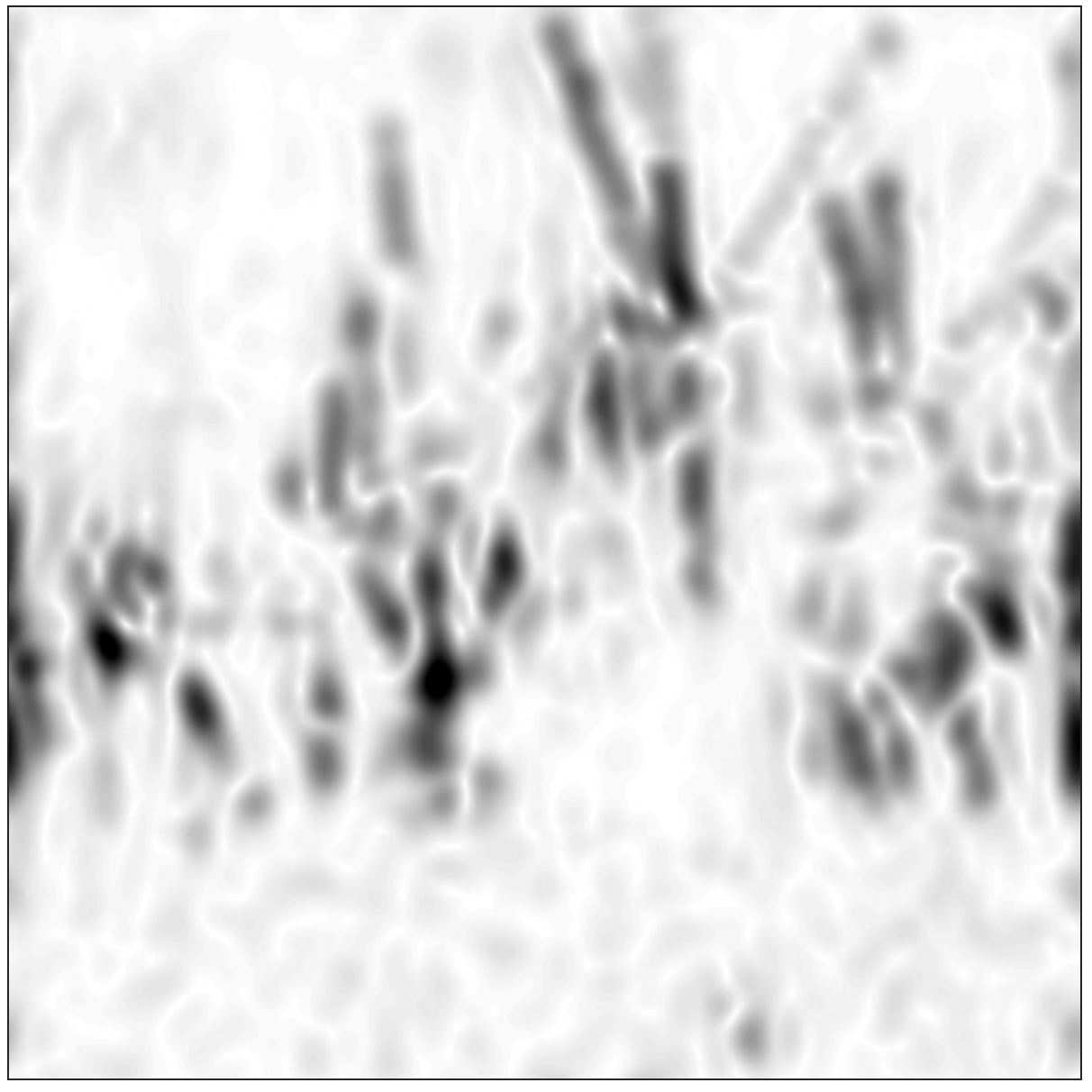}
        \includegraphics[width=1.2in]{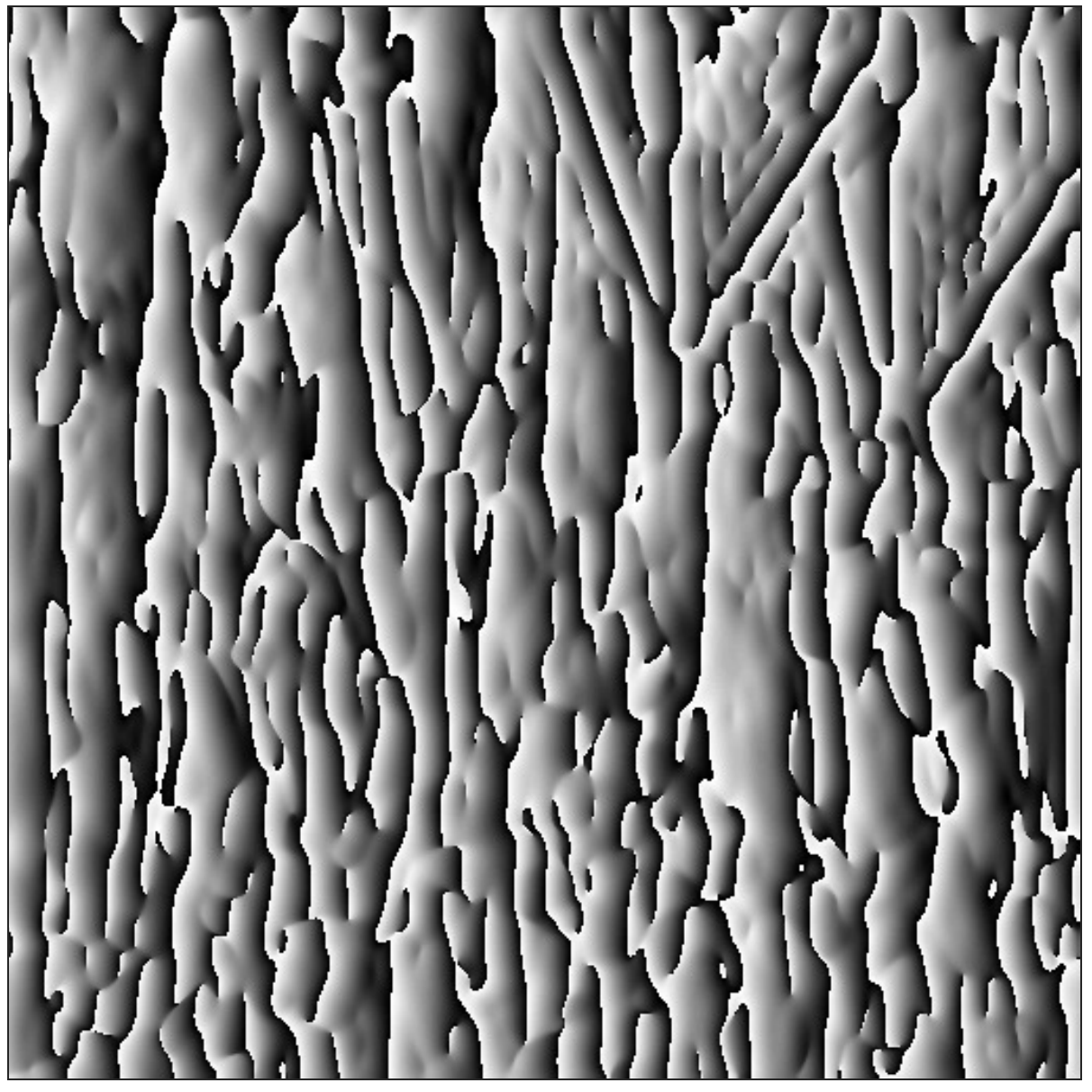}
        \includegraphics[width=1.2in]{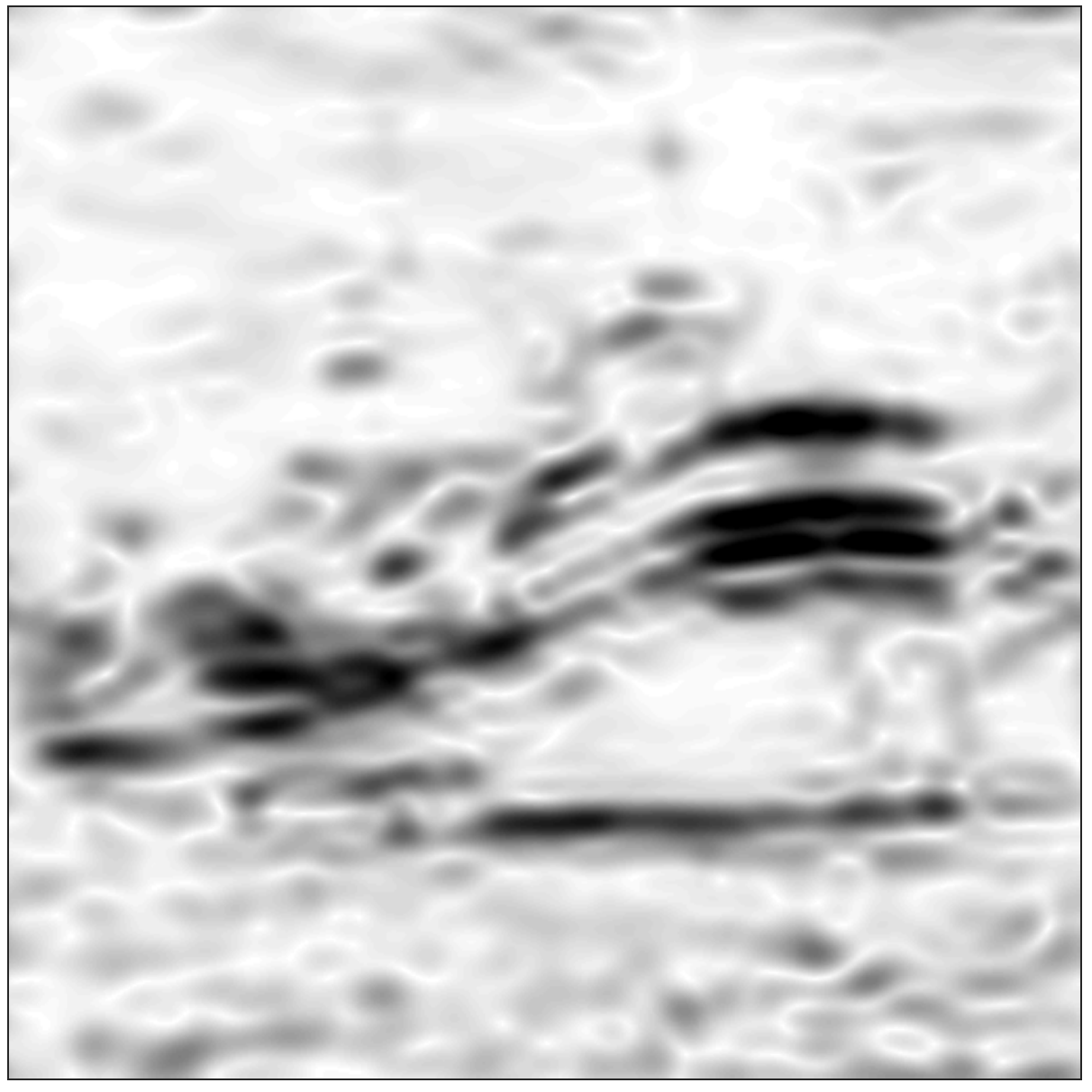}
        \includegraphics[width=1.2in]{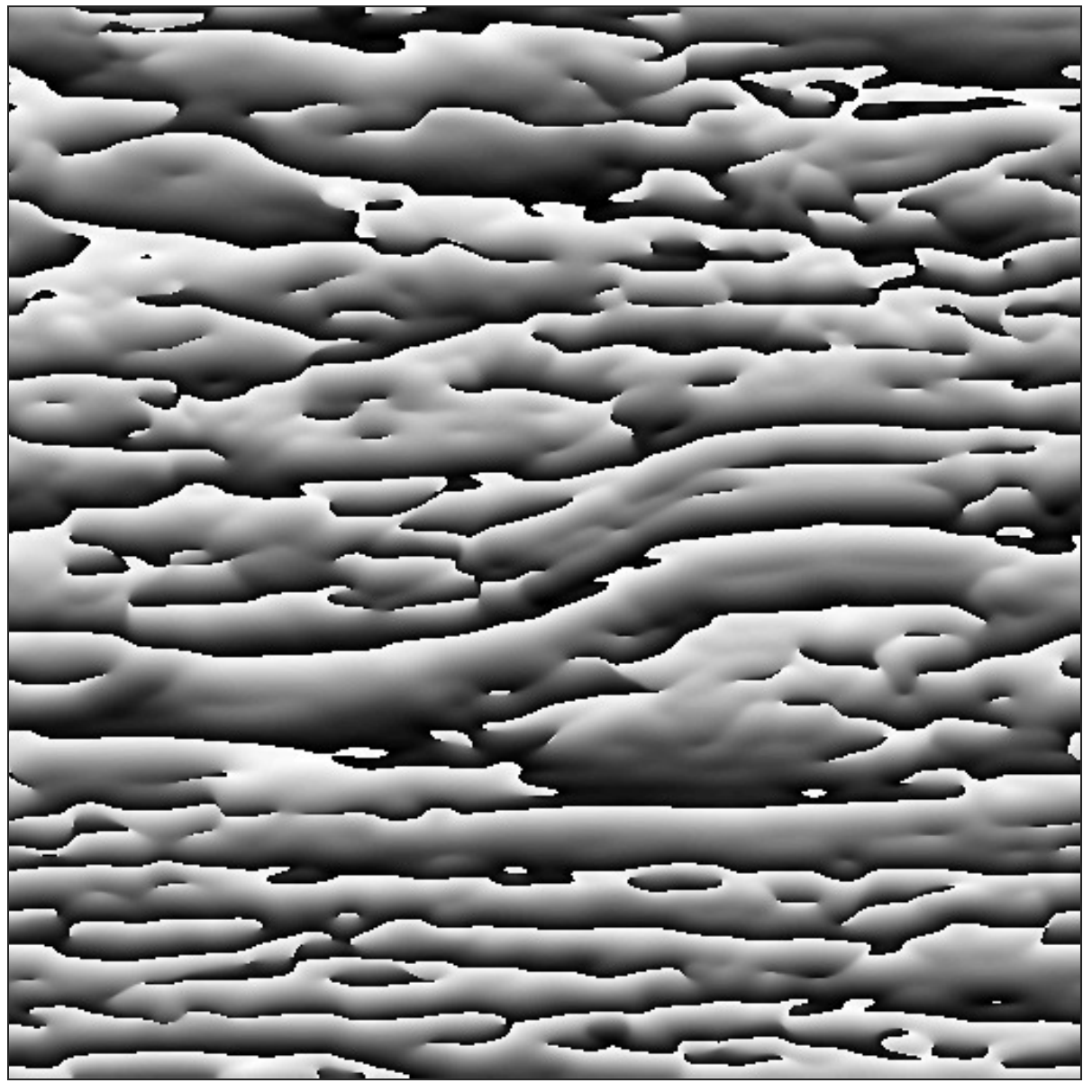}\\
(a)~~~~~~~~~~~~~~~~~~~(b)~~~~~~~~~~~~~~~~~~~(c)~~~~~~~~~~~~~~~~~~~(d)
\end{center}
\caption{Top: original image. Columns (a,c) display
$|x \star \psi_{\la}(u)|$ at scales $2^j = 2 , 4 , 8$ from top to bottom.
Small and large coefficients are displayed 
in white and black respectively. 
The wavelet orientation angle is 
$\theta = 0$ for (a) and $\theta = \pi/2$ for (c).
Columns (b,d) display the phase images
$\varphi(x \star \psi_{\la})$, with $\theta = 0$ for (b) and
$\theta = \pi/2$ for (d).}
\label{sig2D_bumpsteer}
\end{figure}

Let us represent frequencies in polar coordinates
$\omega = |\omega|\, (\cos \theta, \sin \theta)$.
Steerable wavelets introduced
in \citep{steerableSimoncelli}  have
a center frequency $\xi = (\xi_0 , 0)$ with $L$ rotation angles, and
a Fourier transform $\widehat \psi(\om)$ which can be written:
\begin{equation}
\label{bump3}
\widehat \psi(\om) = c\, g\Big(\frac{|\om| - \xi_0}{\xi_0}\Big)\,  \cos^{L-1} (\theta) 1_{ | \theta | < \frac{\pi}{2} } .
\end{equation}
Different wavelets are obtained by modifying the window $g$
\citep{UnserSteerable}.
Steerable bump wavelets are computed with 
the one-dimensional bump window $g$ in (\ref{bump}).
Since $\widehat \psi$ has a compact support and
$L-1$ bounded derivatives,
$\psi$ is 
a ${\bf C}^{\infty}$ function and $|\psi(u)| = O((1 + |u|)^{-L+1})$. 
All partial derivatives of $\widehat \psi(\om)$ are zero at $\om = 0$ so
$\psi$ has an infinite number of vanishing moments
\[
\forall (k_1,k_2) \in \Z^2~~,~~\int_{\R^2} u_1^{k_1}\, u_2^{k_2}\, \psi(u_1,u_2)\, du_1 d u_2 = 0~.
\]
To minimize the Littlewood-Paley constant
$\eta$ in (\ref{littndoisdf}), the constants in (\ref{gaussian2}) and 
(\ref{bump3}) are chosen to be
\[
\sigma_J = 2^{{-0.550}}  \, 2^{-J+1}\,\xi~~\mbox{and}~~
c = 1.29^{-1}\, 2^{L - 1 }  \frac{  (L -1)! } { \sqrt{ {L} [ 2 (L-1)  ] !}  } .
\] 
In numerical applications, we choose $\xi = 0.85\,\pi$ and $L \geq 4$.
One can verify that 
the Littlewood-Paley constant is $\eta = 0.091$ as for the one-dimensional
wavelet (\ref{bump}) and does not depend upon $L$.
Figure \ref{bumpwave2} shows the real and imaginary parts of $\psi$ as
well as its Fourier transform for $L = 8$ angles.

Figure \ref{sig2D_bumpsteer} displays the phase and modulus of the
wavelet transform of an image, calculated with bump steerable wavelets
at angles $\theta = 0$ and $\theta = \pi/2$ and scales $2^j =2,4,8$. 
Wavelet coefficients are sparse at fine scales. 
Large modulus coefficients $|x \star \psi_\la (u)|$
are along edges and sharp transitions. The decay of these modulus values
when $|\la|$ increases depends upon the Lipschitz regularity of $x$ at $u$
\citep{jaffard}. The phase
$\varphi(x \star \psi_\la (u))$ measures 
the local symmetry of the variations of $x \star \psi_\la (u)$ 
when $u$ moves along the direction of $\theta$. 
This symmetry is typically preserved along edges. This is why lines
of constant phase in Figure \ref{sig2D_bumpsteer}
follow the geometry of edges when the modulus
is non-zero. This observation is at the
basis of many edge detection algorithms \citep{mallatbook}.
The phase evolution 
across scales specifies the edge profile.

\section{Phase Dependence Across Frequencies}
\label{invarian}     
This section characterises phase dependencies across frequencies,
from the autocorrelation of $\UU x(u,\la,\alpha)$.
Section \ref{autocnsdfusd} 
introduces the autocorrelation and covariance matrices across
$\lambda$ and $\alpha$ integrated over $u$. 
Section \ref{phaseharmon} proves Lipschitz continuity
properties of the autocorrelation.

\subsection{Phase Harmonic Autocorrelation}
\label{autocnsdfusd}

Section \ref{linsdfsa0} explains that 
$x \star \psi_\la$ is not correlated with $x \star \psi_{\la'}$ if
the  support of 
$\widehat\psi_\la$ and $\widehat \psi_{\la'}$ do not overlap. 
The phase filtering introduces correlations
between such coefficients which allows us to 
characterise phase dependencies with autocorrelation matrices.

\paragraph{Mean vector}
We compute the mean of $\UU x (u,\la,\alpha)$ along $u$ for
each $(\la, \alpha)$.
The mean vector $\In x$ is defined by
\begin{eqnarray}
\label{corr8}
\In x(\la,\alpha) &=& \int \UU x(u,\la,\alpha)\,du \\
&=& \int 
|x \star  \psi_\la(u)|\, h(\alpha-\varphi(x \star  \psi_\la(u))) \, du.
\end{eqnarray}

Since the Fourier transform of $\UU x (u,\la,\alpha)$ along $\alpha$ is
$\widehat \UU x(u,\la,k) = {\widehat {\hh}(k)}\,
[x \star \psi_\la(u)]^{-k}$, the Fourier mean vector is:
\begin{equation}
\label{Adef}
\widehat \In x (\la,k) = \int \widehat \UU x(u,\la,k) 
\,du =  {\widehat {\hh}(k)}\, 
\int [x \star \psi_\la(u)]^{-k} \,du.
\end{equation}
For $k = 0$, $[x \star \psi_\la]^0 = |x \star \psi_\la|$ so 
\[
\widehat \In x(\la,0) = \widehat \hh(0)\,\|x \star \psi_\la \|_1 .
\]
Let us show that $\widehat \In x(\la,k) \ll \widehat \In x(\la,0)$
for $k \neq 0$.
Since $\widehat \In x (\la,k)$ is the Fourier transform
of $\In (\alpha,\la)$ along $\alpha$, 
this implies that
$\In (\alpha,\la)$ remains nearly constant when $\alpha$ varies.

The integral of $[x \star \psi_\la]^k$
is the value of its Fourier transform 
at the zero frequency. If $k = 1$ and $\la \neq 0$ then
$\widehat \In x(1,\la) = 0$ because
$\widehat \psi_\la (0) = 0$.  For $k > 1$ the support of
$[x \star \psi_\la]^k$ is approximately a dilation of the support
of $x \star \psi_\la$ so its Fourier transform remains negligible at $\om = 0$.
This is illustrated by Figure \ref{Fourierradius} for a one-dimensional 
wavelet filter $\psi_\la$.
For one and two-dimensional bump wavelets,
numerical computations show that for all $k \geq 1$
\[
{|\widehat \In x (\la,k)|} \leq \epsilon\, {\widehat \In x(\la,0)} ,
\]
with $\epsilon = 0.017$ for 
the one-dimensional signal of Figure \ref{sig1D} and
$\epsilon = 0.019$ for the image in Figure \ref{sig2D_bumpsteer}.

\begin{figure}
\begin{center}
        \includegraphics[width=5.5in]{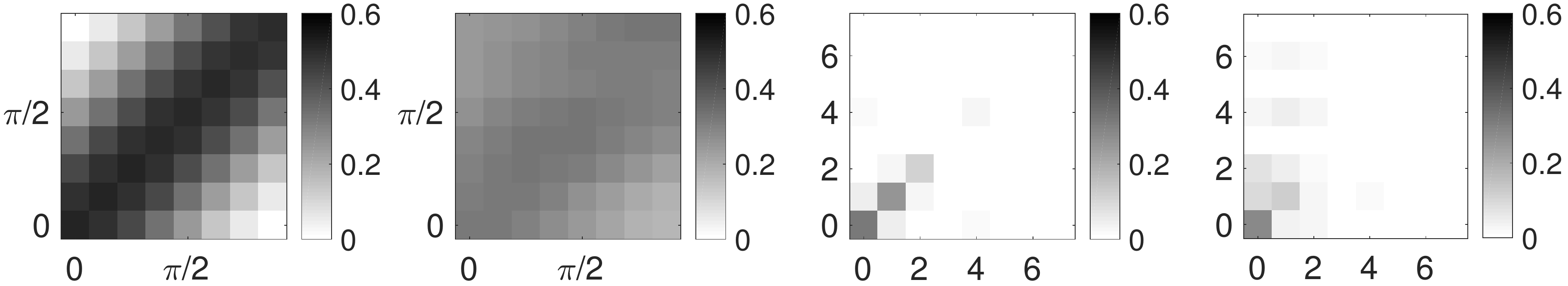} \\
        \includegraphics[width=5.5in]{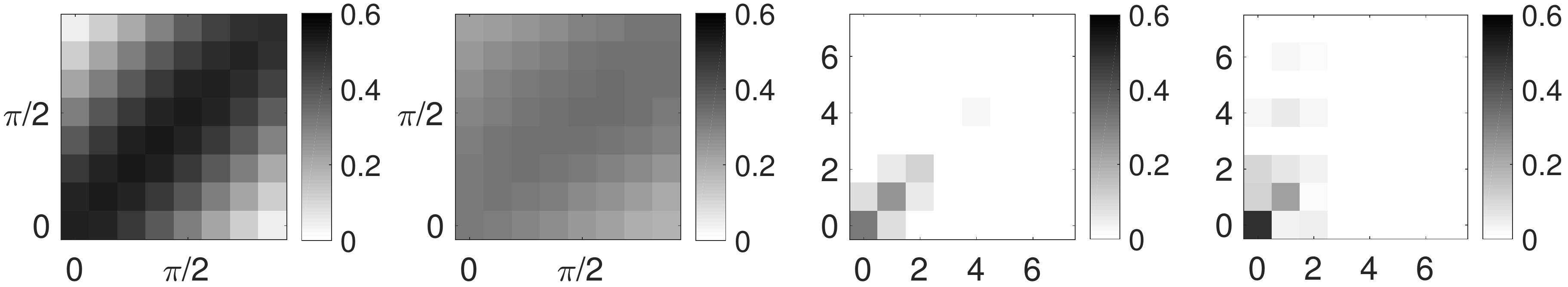} \\
(a)~~~~~~~~~~~~~~~~~~~~~(b)~~~~~~~~~~~~~~~~~~~~~(c)~~~~~~~~~~~~~~~~~~~~~(d)
\end{center}
\caption{Autocorrelation matrices $C x$ and $\widehat C x$
for the one-dimensional signal of Figure \ref{sig1D}  (first row),
and for the image of Figure \ref{sig2D_bumpsteer} (second row).
(a,b): Matrices of $8^2$ coefficients
$ |C x(\la,\alpha,\la',\alpha')|^{1/2}/(\|x \star \psi_\la \|\,\|x \star \psi_{\la'} \|)^{1/2}$ for $(\alpha,\alpha') \in [0,\pi]^2$, with
$\la' = \la$ in (a) and $\la' = 4 \la$ in (b).
(c,d): Matrices of $8^2$ coefficients
$|\widehat C x(\la,k,\la',k')|^{1/2}/(\|x \star \psi_\la \|\,\|x \star \psi_{\la'} \|)^{1/2}$ for $(k,k') \in [0,7]^2$ and 
$\la' = \la$ in (c) and $\la' = 4 \la$ in (d).}
\label{corrl1} 
\end{figure}

\paragraph{Autocorrelation and covariance}
The correlation of $Ux(u,\la,\alpha)$ and 
$Ux(u,\la',\alpha')$ is computed on average along $u$:
\begin{eqnarray}
\label{corr89}
C x(\la,\alpha,\la',\alpha') &=& 
\int  \UU x(u,\la,\alpha)\, \UU x(u,\la',\alpha')^*\, du \\
\label{corr82}
&=& 
\int  |x \star \psi_\la(u)|\, |x \star \psi_{\la'}(u)|\, \\
\nonumber
& &
\nonumber
\hh (\alpha-\varphi(x \star \psi_\la (u))) \, 
\hh( \alpha' -\varphi(x \star \psi_{\la'} (u)))^* \,du .
\end{eqnarray}
It gives the correlation
of $x \star \psi_\la (u)$ and 
$x \star \psi_{\la'} (u)$ at positions $u$ where their
phases are respectively in the
neighborhoods of $\alpha$ and $\alpha'$.
For a fixed $(\la,\la')$, it is a full matrix which has a 
regular $2 \pi$ periodic
oscillation along $(\alpha,\alpha')$.
This is shown in Figure \ref{corrl1} for a one-dimensional signal
and an image.

The Fourier transform along $\alpha$ is an orthogonal change of basis.
We show that it decorrelates a large portion of the coefficients and yields
a sparse autocorrelation matrix. The resulting 
harmonics autocorrelation matrix is
\begin{eqnarray}
\nonumber
\widehat Cx(\la,k,\la',k') &=& \int
\widehat \UU x(u,\la,k) \, \widehat \UU x(u,\la',k')^*\,du\\
\label{Adef9}
& =& {\widehat {\hh}(k)\,\widehat {\hh}(k')^*}\,
\int [x \star \psi_\la(u)]^{-k} \, [x \star \psi_{\la'}(u)]^{k'} \,du.
\end{eqnarray}
Diagonal coefficients are proportional to $\bf L^2$ norms
\begin{equation}
\label{diagonsdfs}
\widehat C x(\la,k,\la,k) = |\widehat \hh(k)|^2\, \|x \star \psi_\la \|^2 .
\end{equation}
The Plancherel formula applied to (\ref{Adef9}) proves that
$\widehat C x(\la,k,\la',k')$ is nearly zero if the Fourier transforms of
$[x \star \psi_\la(u)]^k$ and $[x \star \psi_{\la'}(u)]^{-k'}$ are
concentrated over frequency domains which
do not intersect. 
Since their center frequencies
are approximately $k \la$ and $k' \la'$, the coefficients are negligible
when $|k \la - k' \la'|$ is too large.
The non-negligible coefficients for $k \la \approx k' \la'$  
capture non-linear correlations across frequency bands.

For wavelet filters, numerical experiments show that the Fourier transform of 
$[x \star \psi_\la]^k$ has most of its energy concentrated
in a ball centered in $k \la$ of radius $\beta \max(|k|,1)\, |\la|$,
where $\beta$ depends upon $\psi$. These are qualitative results,
illustrated by Figure \ref{Fourierradius},
but we have no mathematical characterisation of $\beta$.
It results that  $\widehat C x(\la,k,\la',k')$ is non-negligible 
if the distance between the center frequencies $k \la$ and $k' \la'$
satisfies
\begin{equation}
\label{zero-cond}
|k \la - k' \la'| \leq \beta\, (\max(|k|,1) \,|\la| + \max(|k'|,1)\, |\la'|) .
\end{equation}

When $(k,k')$ varies for $(\la,\la')$ fixed, 
(\ref{zero-cond}) defines a band of non-negligible coefficients
centered at  $k \la =  k' \la'$. Beyond this band, 
$\widehat C x(\la,k,\la',k')$ is nearly zero, which yields a sparse
matrix. This property is illustrated by Figure \ref{corrl1}
for the one-dimensional signal of Figure \ref{sig1D}
and for the image of Figure \ref{sig2D_bumpsteer}. 
The first two columns of Figure \ref{corrl1} show that
$Cx (\la,\alpha,\la',\alpha')$ 
is a full matrix for $\la = \la'$ and $\la' = 4 \la$
whereas the Fourier transform matrices
$\widehat Cx (\la,k,\la',k')$ are sparse.

The covariance matrix is
\[
K x = Cx - Mx\, Mx^*~,
\]
where $V^*$ is the complex transpose of a vector $V$.
This covariance measures the dependence of phases 
across frequencies at a same position $u$. 
In the Fourier phase basis,
the first diagonal coefficients of $\widehat Kx = \widehat C x - \widehat Mx\, \widehat Mx^*$ are 
\[
\widehat K x(\la,0,\la,0) = |\widehat h(0)|^2 \Big(\|x \star \psi_\la \|^2 - \|x \star \psi_\la \|_1^2 \Big) .
\]
They measure the sparsity of $x \star \psi_\la(u)$, which 
is large if $\|x \star \psi_\la \|/ \|x \star \psi_\la \|_1$ is small.

\subsection{Lipschitz Continuity for Spectral Norms}

\label{phaseharmon}

We prove that the mean vector $\In x$ is Lipschitz continuous
and that the autocorrelation matrix $C x$ has a bounded trace
and is Lipschitz continuous for the spectral norm. As a result,
they define stable signal representations.
Section \ref{reconsec} shows that these descriptors can provide accurate
signal approximations.

To establish the result for 
$\In x (\la,\alpha) = \int U x(u,\la,\alpha)\, du$,
we suppose that $x$ has
a support in $[0,L]^\dd$ and that all convolutions are 
computed as periodic
convolutions over this domain so that $x \star \psi_\la \in \Ld ([0,L]^\dd)$.
It introduces a normalization factor $L^d$.

For any $x \in \Ld(\R^\dd)$, 
$C x$  is a positive symmetric matrix whose spectral norm is:
\[
\|C x \|_{2,2} = \sup_{v \neq 0} \frac {\|C x\,  v\|} {\|v\|}~~\mbox{with}~~
\|v \|^2 = \frac 1 {2 \pi} \sum_{\la \in \Lambda} \int_0^{2 \pi} |v(\alpha,\lambda)|^2\, d \alpha~, 
\]
and 
\[
C x\,  v(\lambda,\alpha) = \int_0^{2 \pi} \sum_{\la' \in \Lambda}
C(\la,\alpha,\la',\alpha')\, v(\lambda',\alpha')\, d\alpha' .
\]
The following theorem computes Lipschitz bounds for these norms. 

\begin{theorem}
\label{tracthes}
If the filters $\psi_\la$ satisfy (\ref{littndoisdf}) then 
for all $(x,x') \in \Ld ([0,L]^\dd)^2$ 
\begin{equation}
\label{nsdfsdfanusdf9sdf0}
\| \In x -  \In x'\| \leq \kappa\, (1 + \eta)\, L^{\dd/2} \, \|x - x' \|~,
\end{equation}
where $\kappa$ is defined in (\ref{nsdfsdfa8sdfs8sd}).
For all $x \in \Ld (\R^\dd)$
\begin{equation}
\label{nsdfsdfanusdf9sdf}
(1 - \eta)^2\,\|h\|^2\, \|x\|^2 \leq 
{\rm Trace}(C x) \leq (1 + \eta)^2\, \|h\|^2\, \|x\|^2~,
\end{equation}
and all $(x,x') \in \Ld (\R^\dd)^2$
\begin{equation}
\label{nsdfsdfanusdf9sdf89sd}
\|C x - C x' \|_{2,2} \leq \,\kappa^2\, (1 + \eta)^2\,  \|x - x' \|\, ( \|x\| + \|x'\|).
\end{equation}
\end{theorem}

{\it Proof:} We write the mean $\In x$ in (\ref{corr8}) in a vector form
$\In x  = \int Ux(u,.)\, du$, which gives
\[
\| \In x -  \In x'\|^2 = \,
\Big\| \int ( \UU x(u,.) - \UU x'(u,.) )  \,du \Big\|^2 .
\]
If $y \in \Ld([0,L]^\dd)$ 
then the Cauchy-Schwartz inequality implies 
\[
\Big| \int  y(u) du  \Big|^2 \leq L^\dd \int |y(u)|^2 du . 
\]
Applied to the vector $\UU x(u,.) - \UU x'(u,.)$ it gives
\[
\| \In x -  \In x'\|^2 \leq L^\dd \int \| \UU x(u,.) - \UU x'(u,.)\|^2  \,du.
\]
Applying the Lipschitz upper bound (\ref{nsdfsdfanusdfu9sdf})
proves (\ref{nsdfsdfanusdf9sdf0}).

The trace of $C x$ is computed from its diagonal coefficients
in (\ref{diagonsdfs})
\[
\widehat C x(\la,k,\la,k) = |\widehat \hh(k)|^2\, \|x \star \psi_\la \|^2 ,
\]
and hence
\[
{\rm Trace}(C x) = \sum_k |\widehat \hh (k)|^2\, \sum_{\la \in \Lambda} 
\|x \star \psi_\la \|^2  = \|h \|^2\, \|W x \|^2~.
\]
Applying (\ref{nsdfsdf89sdf}) for $x' = 0$ proves (\ref{nsdfsdfanusdf9sdf}).

The operator $Cx$ is defined in (\ref{corr89}) as 
the autocorrelation of $Ux$ which we write in a vector form
$C x = \int Ux(u,.) \, Ux(u,.)^{t} du$, where
$Ux(u,.)^{t}$ is the transpose of $Ux(u,.)$.
We thus verify that
\begin{eqnarray}
\nonumber
Cx' - Cx = \int &\Big(& Ux'(u,.) (Ux'(u,.) - Ux (u,.))^{t} + \\
& & (Ux'(u,.) - Ux (u,.))\,Ux(u,.)^t\Big) du
\label{sni8sdfxc}
\end{eqnarray}
Any matrix $C(a,b) = \int A (u,a)\, B(u,b)\, du$ has
a norm which satisfies
\begin{equation}
\label{sni8sdfxc89sd}
\|C\|_{2,2} \leq \|A \|\, \|B\|
\end{equation}
with $\|A\|^2 = \int \sum_a |A(u,a)|^2 du$ and $\|B\|^2 = \int \sum_b |B(u,b)|^2 du$. This is verified with the Cauchy Schwartz inequality by showing that for any vectors $w$ and $w'$:
\[
|\lb w , C w' \rb| \leq \|w \|\, \|w'\|\, \|A \|\, \|B\|~.
\]
Applying (\ref{sni8sdfxc89sd}) to (\ref{sni8sdfxc}) proves that
\[
\|C x' - C x \|_{2,2} \leq  \|\UU x' - \UU x\| \, ( \|\UU x'\| + \| \UU x\|) .
\]
Since $\|\UU x - \UU x'\| \leq \kappa\, (1 + \eta)\, \|x - x' \|$ and
$\|\UU x \| \leq \|h\|\, (1 + \eta)\, \|x\|$ with 
$\|h \|^2 \leq \kappa^2$, we derive (\ref{nsdfsdfanusdf9sdf89sd}). $\Box$

The matrix $C x$ is a positive symmetric operator so
$\|C x\|_{2,2} \leq {\rm Trace}(C x)$. The theorem proves 
in (\ref{nsdfsdfanusdf9sdf}) that $C x$ is a bounded matrix.
It also proves in (\ref{nsdfsdfanusdf9sdf89sd}) 
that $C x$ is Lipschitz continuous when $x$ 
varies in the neighborhood of $x'$. 
The theorem results also apply to $\widehat \In x$ and $\widehat Cx$
which are obtained from $\In x$ and $C x$ through a Fourier
orthogonal change of basis.

\section{Compressive Approximations from Harmonic Correlations}
\label{reconsec}

Autocorrelations of one layer neural network coefficients 
have been studied 
in \citep{Bethdge2} to generate stationary image textures 
having similar perceptual properties as an original texture $x$. 
In this section we study the reconstruction of $x$ up to 
a global translation, as opposed to a perceptually similar signal.

To reconstruct $x(u)$ from its autocorrelation when
shifting $x$ along $u$ is equivalent to recover $x$ from 
its Fourier transform modulus $|\widehat x|^2$. This is only
possible up to a global translation. It amounts
to solve a phase retrieval problem which has been widely studied
from mathematical \citep{FourierRecons} and algorithmic point of views
\citep{phaserecovery}. Several authors have shown that this recovery
from autocorrelations
can be solved with a reduced number of autocorrelation
measurements if $x$ is sparse 
\citep{sparsephaserecovery1,sparsephaserecovery2}.
The recovery is performed by minimizing an $\bf l^1$ norm to enforce sparsity.

Next sections study the recovery of $x$ from the mean and autocorrelation
of $U x (u,\la,\alpha)$. The autocorrelation is not computed with a
shift along $u$ but along $(\la,\alpha)$. It relies on the 
phase correlation
created by harmonics but it is mathematically more 
complicated because $U$ is non-linear.
We introduce a recovery algorithm based on 
a gradient descent, which also takes advantage of sparsity through
$\bf l^1$ norm conditions.
Section \ref{numericsec} shows that nearly optimal approximations
can be obtained from wavelet harmonic correlations, if the signal has
sparse wavelet coefficients. Computations are reproduced by a software 
in {\it https://github.com/kymatio/phaseharmonics}.

\subsection{Compressive Recovery}

We introduce a compressive recovery
algorithm which computes a signal approximation
from a limited set of $\M$ harmonic means and
correlations. These means and correlations are invariant to translations
so the recovery is up to a global translation.

Computations are carried over 
$\dd$-dimensional signals $x(u)$ uniformly
sampled over $N^\dd$ points, with $\dd = 2$ for images.
We set $\widehat h(k) = 1$ over a limited range of $k$ and 
$\hat h(k) = 0$ beyond.
The mean vector is computed with sums which are normalized by $N^\dd$
\begin{equation}
\label{spansdfc0sdfs1}
\widehat \In x(\la,k) = N^{-\dd} \sum_{u} [x \star \psi_\la (u)]^k~.
\end{equation}
Section \ref{autocnsdfusd} shows that $\widehat \In x(\la,k)$ is
non-negligible only for $k = 0$
\begin{equation}
\label{spans1l1}
\widehat \In x(\la,0) = N^{-\dd} \|x \star \psi_\la \|_1 .
\end{equation}
This $\bf l^1$ norm specifies the sparsity of $x \star \psi_\la (u)$.

The autocorrelation of $\widehat \UU x$ is
\begin{equation}
\label{spansdfc0sdfs}
\widehat C x(\la,k,\la',k') = N^{-\dd} \sum_{u} [x \star \psi_\la (u)]^{-k}\,[x \star \psi_{\la'} (u)]^{k'} .
\end{equation}
Corollary \ref{recnsdf} proves that $x(u)$ is recovered from 
$\widehat \UU x(u,\la,k)$
with a linear inverse operator. 
Recovering $\widehat \UU x$ from
a limited number of autocorrelation coefficients $\widehat C x$
and means $\widehat M x$ can be interpreted
as a quadratic recovery problem conditioned by
$\bf l^1$ norm sparsity constraints.
However, $\widehat \UU x(u,\la,k)$ is also correlated along $u$
because the filters $\psi_\la (u)$ are regular. 
Instead of trying to recover $\widehat \UU x$ having such a regularity,
we recover $x$ directly.

\paragraph{Loss minimization}
We want to recover a signal $\tilde x$ such that
$\widehat M \tilde x$ and $\widehat C \tilde x$ are equal to 
$\widehat Mx$ and $\widehat Cx$ over a predefined subset of
$\M$ coefficients. It is equivalent to match $\widehat Mx$ and the
covariance $\widehat Kx = \widehat Cx - \widehat Mx\,\widehat Mx^*$.
The loss ${\cal E}_\M (x,y)$ of an approximation $y$ of $x$
is defined from a discrepancy between the means and covariances of
$\widehat U y$ and $\widehat U x$.
In a matrix form it is defined by
\begin{equation}
\label{errnsdfoisd}
{\cal E}_\M (x,y) = 
\| \widehat Ky - \widehat Kx + (\widehat \In y - \widehat \In x)\, (\widehat \In y - \widehat \In x)^* 
\|_F^2 ,
\end{equation}
where
$\| A \|_F$ is the Frobenius norm computed over
a selected set of $\M$ indices $(\la,k,\la',k')$.

We want to find $y = \tilde x_\M$ which minimizes ${\cal E}_\M (x,y)$.
This is done with a gradient descent initialized with 
a Gaussian white noise $y_0$. The
gradient descent computes $y_{n+1}$ from 
$y_{n}$ with a gradient step on ${\cal E}_\M (x,y)$ at $y_n$.  
We use an unconstrained gradient descent algorithm L-BFGS.
The algorithm stops the minimization with the line-search Wolfe condition
\citep{OptimBook}. The loss is not convex so
the gradient descent may be trapped in local minima.
To improve local minima,
we compute the gradient descent with $10$ random initializations 
and we keep the solution $\tilde x_\M = y_n$ having a minimum loss
${\cal E}_\M (x,y_n)$.

\subsection{Recovery From Wavelet Harmonic Correlations}
\label{numericsec}

We evaluate approximation errors of one and two-dimensional signals
from $\M$ wavelet harmonic autocorrelations and means.
We demonstrate numerically that this representation has 
compressive approximation properties. 
It recovers an accurate approximation
of signals having sparse wavelet representations from fixed
sets of wavelet harmonic correlations.
The error decay rate is comparable
to sparse non-linear approximations in a wavelet basis, which requires
to adjust the choice of wavelet coefficients to each signal. 
However, wavelet harmonic correlations do not reconstruct signals
having wavelet coefficients which are not sufficiently sparse. 

\paragraph{Correlation selection}
We first explain how to select 
$\M$ non-negligible wavelet harmonic correlation coefficients,
independently from $x$.

Section \ref{autocnsdfusd} shows that $\widehat Cx(\la,k,\la',k')$ is 
non-negligible only if $k\la$ is sufficiently close to $k' \la'$:
\begin{equation}
\label{zero-cond8}
|k \la - k' \la'| \leq \beta\, (\max(|k|,1) \,|\la| + \max(|k'|,1)\, |\la'|) .
\end{equation}
for some constant $\beta$.
Since $\widehat C x$ is symmetric, we can impose  $|\la| \geq |\la'|$. 
We only keep low-order harmonics
by setting $k = 0$ or $k = 1$, and for each $\la'$ we restrict $k'$ to
the values which satisfy (\ref{zero-cond8}).

Let $Q$ be the number of wavelet scales per octave.
The wavelet frequencies $\la$ and
$\la'$ correspond to scales $2^{j/Q} \sim |\la|^{-1}$ and 
$2^{j'/Q} \sim |\la'|^{-1}$. 
We limit the range $\Delta$ of scale interactions by imposing that
\begin{equation}
\label{range}
|j-j'| \leq \Delta \, Q .
\end{equation}
In dimension $d$, a signal of $N^\dd$ samples has at most
$J \leq \log_2 N$ dyadic scales $2^j$, so $\Delta \leq \log_2 N$.
In all numerical experiments we set $J = \log_2 N$, and all convolutions
are computed with periodic boundary conditionsf.

For one-dimensional signals of size $N$,
the total number $\M$ of mean and correlation coefficients
which satisfy (\ref{zero-cond8}) and (\ref{range}) is
\begin{equation}
\label{Mvalue}
\M \sim  \Delta^2\, Q^2\, \log_2 N ~.
\end{equation}
For images of $N^2$ pixels, we set $Q = 1$ but
the number of coefficients also
depends upon the number $L$ of wavelet angles, and hence
 \begin{equation}
\label{Mvalue2}
\M \sim  \Delta^2\, L^2\, \log_2 N.
\end{equation}

\begin{figure}
\begin{center}
        \includegraphics[width=1.6in]{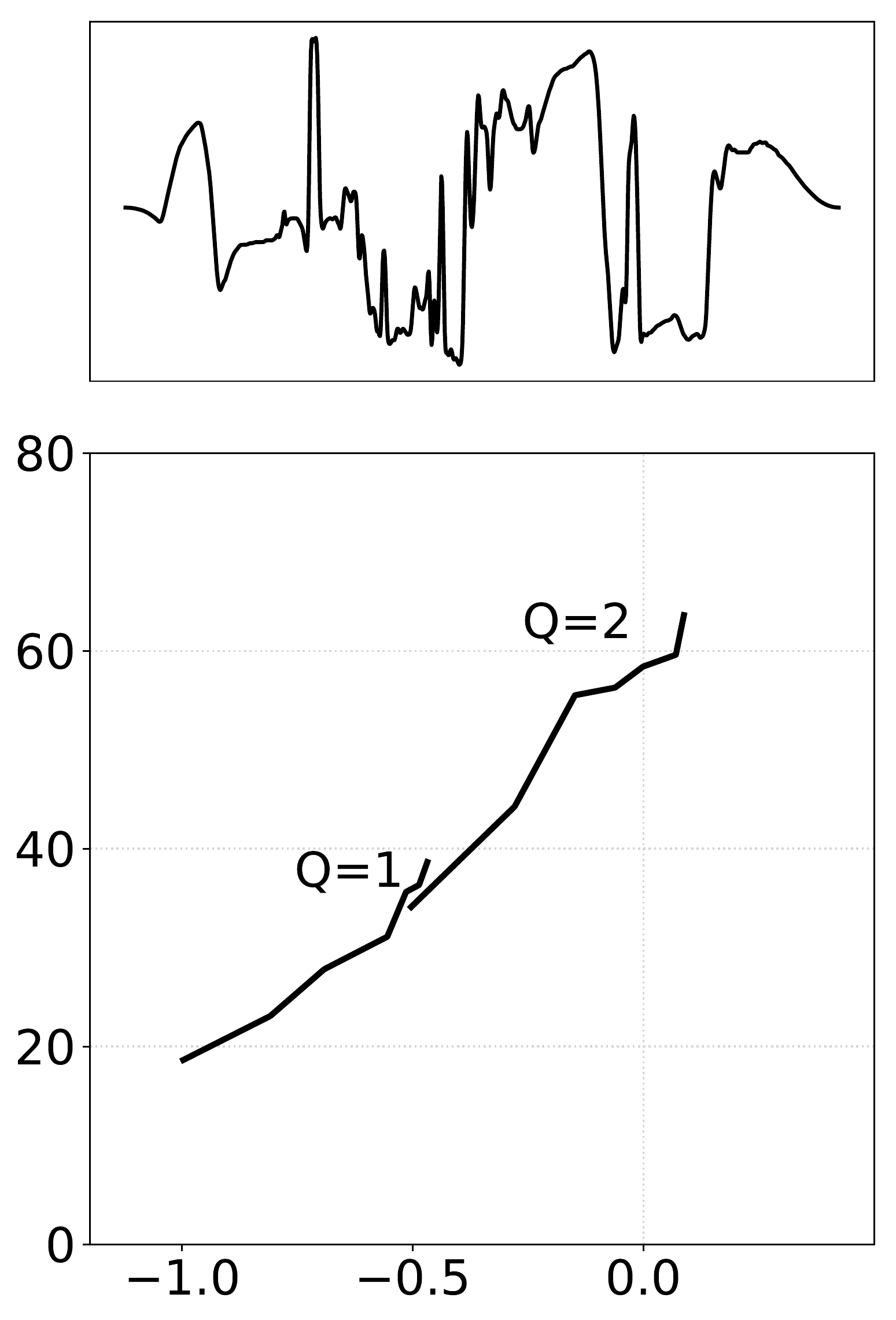}
        \includegraphics[width=1.6in]{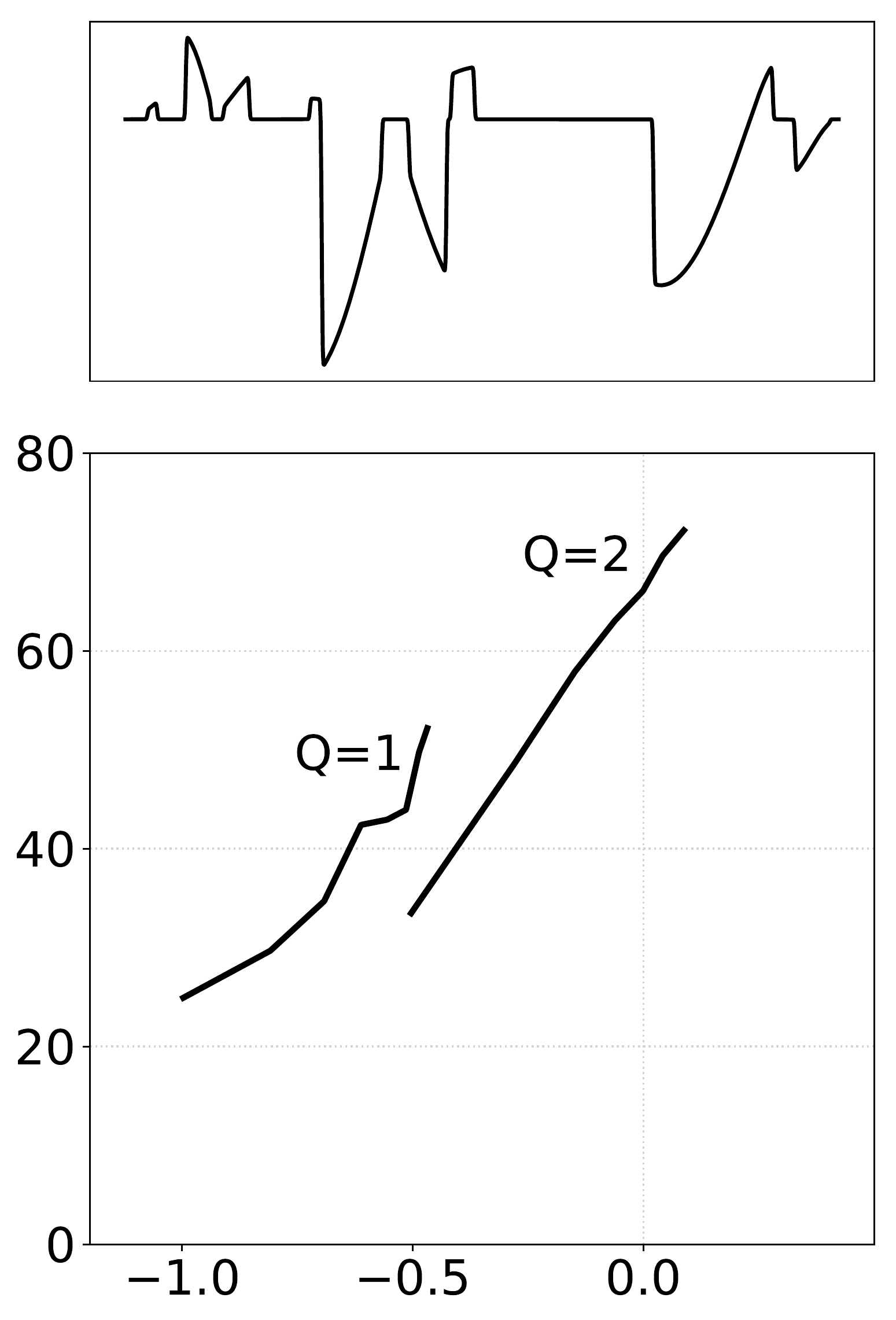}
        \includegraphics[width=1.6in]{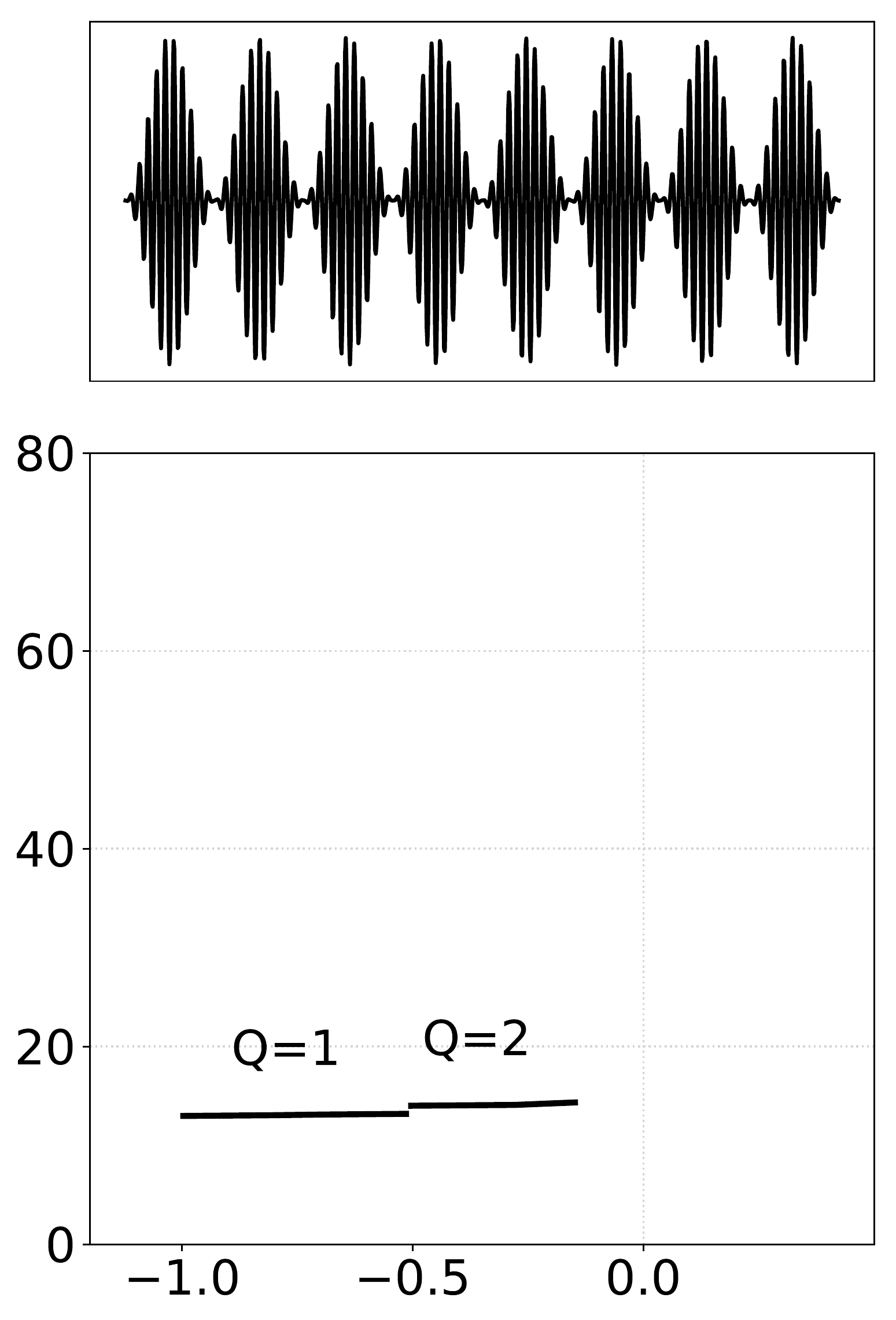}\\
(a)~~~~~~~~~~~~~~~~~~~~~~~~~~~(b)~~~~~~~~~~~~~~~~~~~~~~~~~~~(c)
\end{center}
\caption{For each figure, the top graph is 
the original signal $x$ of $N = 1024$ points.
The curve below gives the PSNR 
of signals $\tilde x$ reconstructed from $\M$ wavelet harmonic
correlations and means, as a function of 
$\log_{10} \M/N$. The first and second parts 
of each curve correspond to $Q = 1$ and $Q = 2$ respectively,
for $\Delta$ varying.}
\label{recons}
\end{figure}

\begin{figure}
\begin{center}
        \includegraphics[width=2in]{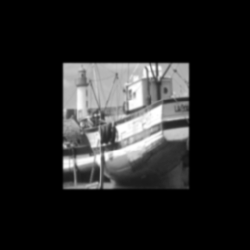}
        \includegraphics[width=2in]{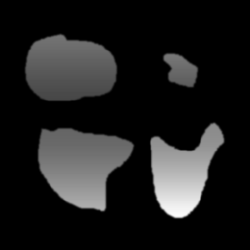}\\
        \includegraphics[width=2.3in]{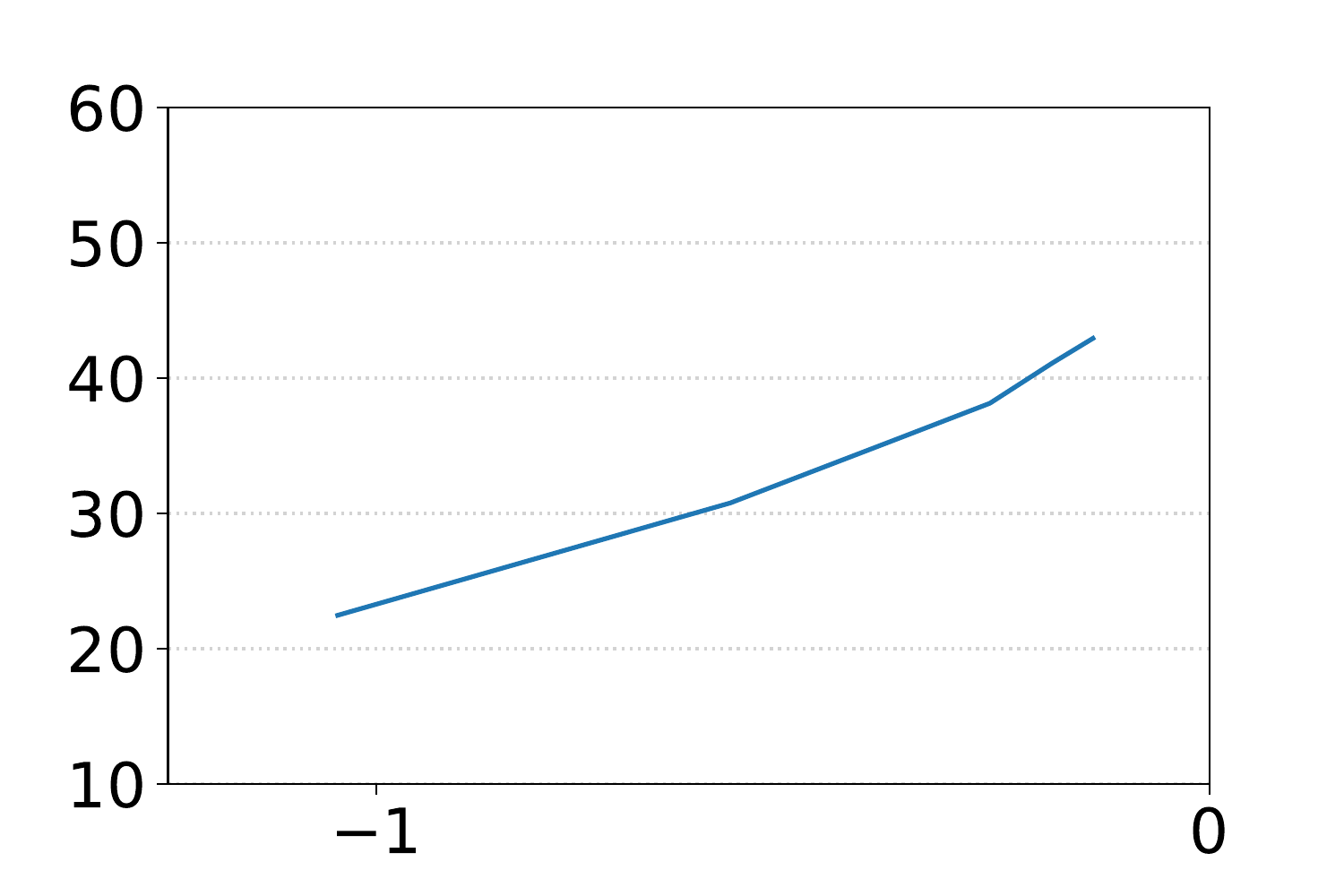}
        \includegraphics[width=2.3in]{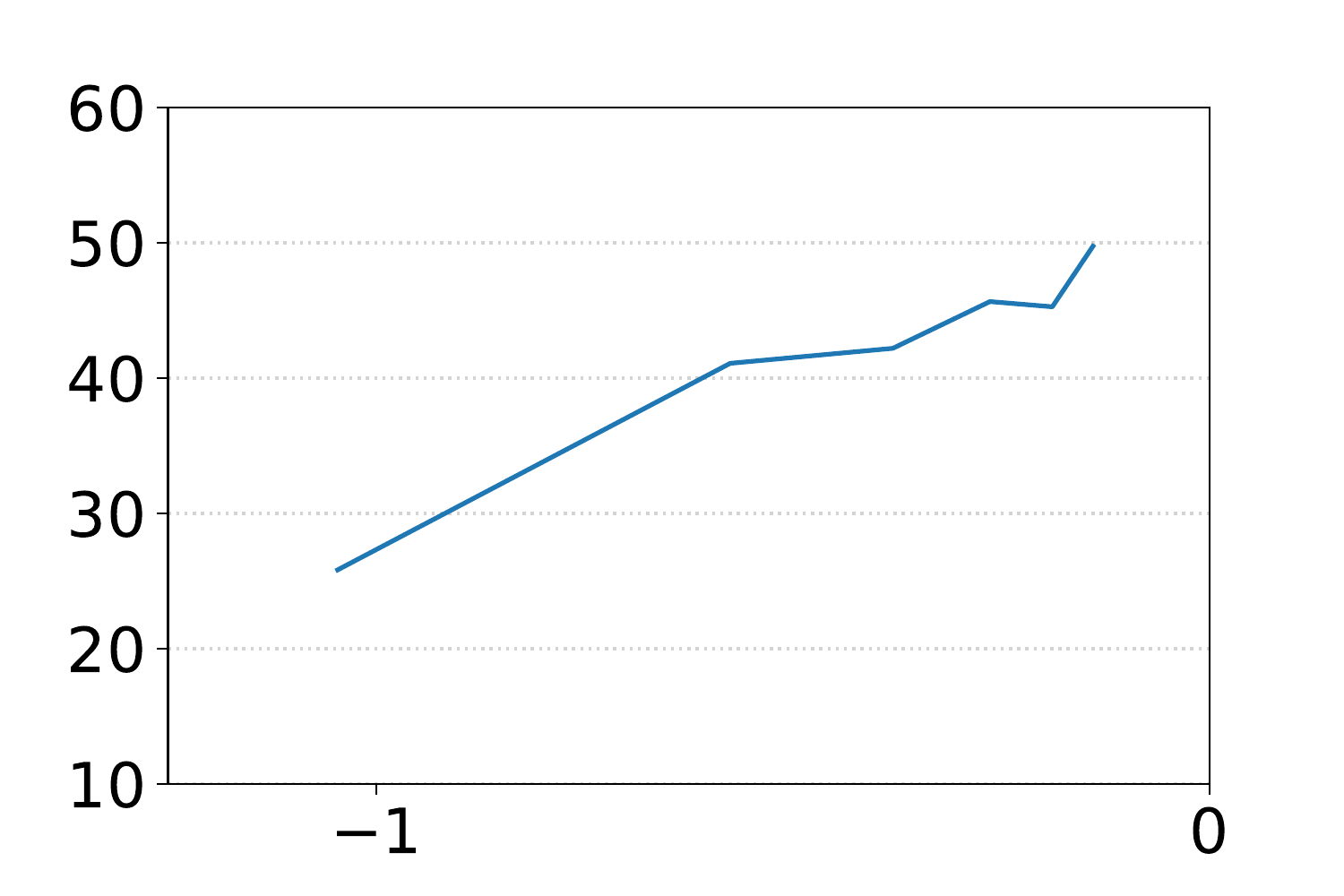}\\
(a)~~~~~~~~~~~~~~~~~~~~~~~~~~~~~~~~~~~~(b)
\end{center}
\caption{Each original image $x$ shown at the top has
$N^2 = 256^2$ pixels. The curve below
gives the PSNR of signals $\tilde x$ reconstructed from 
$\M$ wavelet harmonic correlations and mean coefficients,
as a function of $\log_{10} \M/N^2$. }
\label{recons2}
\end{figure}

\paragraph{Compressive recovery}
Since $\widehat C x$ and $\widehat \In x$ are invariant to translations,
$x$ can only be recovered up to a translation.
The approximation error is calculated by 
translating $\tilde x_\M$ so that it minimizes
$\|x - \tilde x_\M \|$.
We evaluate numerical reconstruction errors from $\M$ 
correlation invariants with
the Peak Signal to Noise Ratio (PSNR) in dB
\[
{\rm PSNR}(x,\tilde x_\M) = 20 \log_{10} \frac {N^{\dd/2}\, \max_u |x(u)|} {\|x - \tilde x_\M\|}~,
\]
Above $35 {\rm dB}$, reconstructed images are visually identical to the original ones, and signal plots superimpose so we do not display reconstructed signals. 
Figure \ref{recons} 
gives the PSNR error 
as a function of $\log_{10} \M/N$ for one dimensional signals,
and Figure \ref{recons2} 
as a function of $\log_{10} \M/N^2$ for images.
The number of coefficients $\M$ varies by adjusting the range $\Delta$
of scale interactions in (\ref{Mvalue}) and (\ref{Mvalue2}).

Figure \ref{recons}(a,b) gives two examples of piecewise 
regular one-dimensional signals
$x$ having sparse wavelet coefficients. Large coefficients are
located at sharp transitions.
The PSNR curves show that when $\M = N$, $\tilde x_\M$ is above $60 {\rm dB}$
which corresponds to a relative error below $10^{-3}$.
The approximation error has a decay
\begin{equation}
\label{errnwsdf}
\| x  - \tilde x_\M\| \leq C\,  \M^{-\chi} ,
\end{equation}
with $\chi \approx 2$. This is the same 
approximation rate as the one obtained by a
non-linear adaptive approximation
of such signals in a wavelet orthonormal basis. 
Non-linear approximations take advantage of sparsity by
selecting the $\M$ largest wavelet coefficients of $x$, whose locations
thus depend upon $x$ \citep{mallatbook}. The decay rate $\chi = 2$ is
the best one obtained over the class of signals 
which may be discontinuous but have a bounded total variation.
In this case, the same error decay rate is obtained over
fixed sets of $\M$ wavelet harmonic correlations, which do not
depend upon $x$ as in adaptive approximations. 
Correlations are averaged over all spatial positions
which is why the choice of correlation 
coefficients is not adapted to the positions
of non-zero wavelet coefficients.
These non-linear approximation properties
over fixed
sets of measurements are similar to the ones
obtained by compressive sensing algorithms
\citep{candes}. However, the mathematical setting is more
difficult because the linear sensing operator is
replaced by non-linear harmonic autocorrelations.

Similarly to compressive sensing, if wavelet coefficients are not
sufficiently sparse then the signal is 
not reconstructed from wavelet harmonic correlations. 
The signal in Figure \ref{recons}(c) is a dramatic example:
\begin{equation}
\label{cossig}
x(u) = (1 - \cos(\nu u))\, \cos (\la u)~~\mbox{with}~~\nu \ll \lambda~.
\end{equation}
The reconstruction algorithm is unable to recover an approximation even
when the number $\M$ of correlation coefficients reaches the signal
size $N$. In this case
the cosine of high frequency $\la$ creates many non-zero wavelet
coefficients $x \star \psi_\la(u)$.
The cosine amplitude is modulated by another
cosine of much lower frequency $\nu$. The phase dependence of
such frequencies need to be captured by harmonic coefficients
providing the correlation of $x \star \psi_\la$ with 
$[x \star \psi_{\nu}]^{k}$ for $k = \la/\nu$. 
However we cannot recover the phase dependence of the two
frequency components because $x \star \psi_\nu = 0$.
When signals have localized sharp transitions, large wavelet
coefficients propagate across scales as shown 
in Figure \ref{sig1D}. At a location $u$, if 
$x \star \psi_{\la'}(u)$ and $x \star \psi_{\la}(u)$ are not zero then 
harmonic correlation coefficients capture the phase dependencies
and can thus reconstruct the signal. 
This is the case for the two signals in 
Figure \ref{recons}(a,b).

Figure \ref{recons2} shows results similar to 
Figure \ref{recons}, for images of $N^2$ pixels.  
These images have sparse wavelet coefficients located near edges.
The reconstruction $\tilde x_\M$ has a PSNR well above $40 {\rm dB}$ and hence
a relative precision of $10^{-2}$ when $\M$ is close to $N^2$,
for the natural image and the piecewise regular cartoon image. 
The error $\|x - \tilde x_\M\|$ decays when increasing
the number $\M$ of correlation terms,
with an exponent $\chi \approx 1$ in (\ref{errnwsdf}).
Again, this is the same 
approximation rate as the one obtained by a non-linear 
adaptive approximation in a wavelet orthonormal basis,
where the $\M$ largest wavelet
coefficients are selected depending upon the image.
This decay exponent is the optimal approximation rate for the class of
images having a bounded total variation \citep{mallatbook}. 
However, the choice of the 
$\M$ harmonic wavelet correlations does not depend upon
the image, as in compressive sensing approximation.

Similarly to the one-dimensional case, 
images whose wavelet coefficients are not
sufficiently sparse cannot be 
reconstructed from wavelet phase harmonic correlations.
One may construct such images with two-dimensional
high frequency sinusoidal waves as in the one-dimensional
example (\ref{cossig}). Realizations 
of ergodic stationary processes are other counter examples.

\paragraph{Ergodicity versus compressive approximations}
Realizations of ergodic stationary processes 
cannot be recovered from a limited number of wavelet harmonic
correlations. This observation relates the signal recovery problem to 
approximations of stationary processes \citep{bruna-stoch}
and image texture synthesis \citep{steerableSimoncelli,Bethdge2}.

If $x$ is
a realization of an ergodic stationary process $X$ and the
domain size $N^\dd$ is sufficiently large then the 
spatial empirical
means (\ref{spansdfc0sdfs1}) and correlations (\ref{spansdfc0sdfs})
provide accurate estimations of expected means 
${\mathbb E}\Big(\widehat UX(u,\la,k)\Big)$ and
correlations 
${\mathbb E}\Big(\widehat UX(u,\la,k)\,\widehat UX(u,\la,k)^*\Big)$,
computed relatively to the probability distribution of $X$.
If $x$ and $x'$ are two realizations of $X$ then $\|x - x'\|$ is
typically large but they 
have nearly the same empirical 
means (\ref{spansdfc0sdfs1}) and correlations (\ref{spansdfc0sdfs}).
It results that if $x$ is 
a realization of an ergodic process over a large domain
size $N^d$ then its
recovery from $\widehat \In x$ and $\widehat C x$ is not stable. 

In this unstable ergodic regime, 
a gradient descent on the loss (\ref{errnsdfoisd}) will reconstruct
different signals for different Gaussian white noise initializations.
Such reconstructions are realizations of a random
process whose probability measure was obtained by transporting the
uniform Gaussian white noise measure with the gradient descent on the loss.
One can prove that
it defines a stationary microcanonical process, conditioned by
the empirical means and autocorrelation coefficients $\widehat \In x$
and $\widehat C x$ \citep{bruna-stoch}.
Texture synthesis algorithms from deep convolutional network coefficients
follow this principle \citep{Bethdge1,Bethdge2}.
The properties of the stationary processes obtained with
wavelet phase harmonic correlations are studied in
\citep{ZhangMallat}. 

This analysis shows that we must distinguish two cases.
If $x$ is a signal whose wavelet coefficients are sufficiently sparse
compared to the total size $N^d$ then it may be recovered from a relatively
small number of wavelet harmonic correlations.
On the contrary, we cannot recover $x$ 
if we are in an ergodic regime where $\widehat \In x$
and $\widehat Cx$ are close approximations of empirical means
and correlations of a stationary ergodic random process $X$.
In this case, the reconstruction algorithm initialized over Gaussian white
noise defines a stochastic model of $X$ \citep{ZhangMallat}.

\section*{Acknowledgment}
This work was supported by the ERC InvariantClass 320959.


\begin{thebibliography}{}

\bibitem[Akutowicz, 1956]{FourierRecons}
Akutowicz, E.~J. (1956)  On the determination of the phase of a Fourier
  integral. {\em Trans. of the American Mathematical Society}, \textbf{83}(1),
  179--192.

\bibitem[Bruna \& Mallat, 2019]{bruna-stoch}
Bruna, J. {\&} Mallat, S. (2019)  Multiscale Sparse Microcanonical Models. {\em
  arXiv:1801.02013}.

\bibitem[Candes et~al., 2006]{candes}
Candes, E., Romberg, J. {\&} Tao, T. (2006)  Robust uncertainty principles:
  exact signal reconstruction from highly incomplete frequency information.
  {\em IEEE Transactions on Information Theory}, \textbf{52}(2), 489--509.

\bibitem[Candes et~al., 2013]{phaserecovery}
Candes, E., Strohmer, T. {\&} Voroninski (2013)  Phaselift: Exact and stable
  signal recovery from magnitude measurements via convex programming. {\em
  Communications on Pure and Applied Mathematics}, \textbf{66}, 1241--1271.

\bibitem[Gatys et~al., 2015]{Bethdge1}
Gatys, L., Ecker, A. {\&} Bethge, M. (2015)  Texture Synthesis Using
  Convolutional Neural Networks. In Cortes, C., Lawrence, N.~D., Lee, D.~D.,
  Sugiyama, M. {\&} Garnett, R., editors, {\em Advances in Neural Information
  Processing Systems 28}, pages 262--270. Curran Associates, Inc.

\bibitem[Grossmann et~al., 1989]{Grossmann}
Grossmann, A., Kronland-Martinet, R. {\&} Morlet, J. (1989)  Reading and
  understanding continuous wavelet transforms. In Combes, J., editor, {\em
  Wavelets, time-frequency representations and phase space}. Springer, Berlin.

\bibitem[Jaffard, 1991]{jaffard}
Jaffard, S. (1991)  Pointwise smoothness, two-microlocalisation and wavelet
  coefficients. {\em Publications Matematiques}, \textbf{35}, 155--168.

\bibitem[Jamshidi \& Kirby, 2006]{Bump}
Jamshidi, A. {\&} Kirby, M.~J. (2006)  Examples of Compactly Supported
  Functions for Radial Basis Approximations. In {\em International Conference
  on Machine Learning}.

\bibitem[Krizhevsky et~al., 2012]{Imagenet}
Krizhevsky, A., Sutskever, I. {\&} Hinton, G.~E. (2012)  ImageNet
  Classification with Deep Convolutional Neural Networks. In {\em Proc. of
  NIPS}, pages 1106--1114.

\bibitem[LeCun et~al., 2015]{CNNnature}
LeCun, Y., Bengio, Y. {\&} Hinton, G.~E. (2015)  Deep learning. {\em Nature},
  \textbf{521}(7553), 436--444.

\bibitem[Luo \& Mesgarani, 2018]{Tasnet}
Luo, Y. {\&} Mesgarani, N. (2018)  TaSNet: Time-Domain Audio Separation Network
  for Real-Time, Single-Channel Speech Separation. In {\em Proc. ICASSSP},
  pages 696--700.

\bibitem[Mallat, 2001]{mallatbook}
Mallat, S. (2001) {\em A Wavelet Tour of Signal Processing: The Sparse Way, 3rd
  Edition}.
Academic Press.

\bibitem[Mallat, 2016]{mallatReview}
Mallat, S. (2016)  Understanding Deep Convolutional Networks. {\em Phil. Trans.
  of Royal Society A}, \textbf{374}(2065).

\bibitem[Moravec et~al., 2007]{sparsephaserecovery1}
Moravec, M., Romberg, J. {\&} Baraniuk, R. (2007)  Compressive phase retrieval.
  In {\em SPIE International Symposium on Optical Science and Technology}.

\bibitem[Nocedal \& Wright, 2006]{OptimBook}
Nocedal, J. {\&} Wright, S.~J. (2006) {\em Numerical Optimization}.
Springer, New York, NY, USA, second edition.

\bibitem[Portilla \& Simoncelli, 2000]{Portilla}
Portilla, J. {\&} Simoncelli, E.~P. (2000)  A Parametric Texture Model based on
  Joint Statistics of Complex Wavelet Coefficients. {\em International Journal
  of Computer Vision}, \textbf{40}, 49--71.

\bibitem[Shechtman et~al., 2011]{sparsephaserecovery2}
Shechtman, Y., Eldar, Y.~C., Szameit, A. {\&} Segev, M. (2011)  Sparsity based
  sub-wavelength imaging with partially incoherent light via quadratic
  compressed sensing. {\em Opt. Express}, \textbf{19}, 14808--14822.

\bibitem[Simoncelli \& Freeman, 1995]{steerableSimoncelli}
Simoncelli, E.~P. {\&} Freeman, W.~T. (1995)  The Steerable Pyramid: A Flexible
  Architecture for Multi-Scale Derivative Computation. In {\em IEEE Int. Conf.
  on Image Processing}, pages 444--447.

\bibitem[Unser et~al., 2011]{UnserSteerable}
Unser, M., Chenouard, N. {\&} Van De~Ville, D. (2011)  Steerable Pyramids and
  Tight Wavelet Frames in ${L}^2 ({R}^d)$. {\em {IEEE} Transactions on Image
  Processing}, \textbf{20}(10), 2705--2721.

\bibitem[Ustyuzhaninov et~al., 2017]{Bethdge2}
Ustyuzhaninov, I., Brendel, W., Gatys, L. {\&} Bethge, M. (2017)  What does it
  take to generate natural textures?. In {\em International Conference on
  Learning Representations}.

\bibitem[Zhang \& Mallat, 2019]{ZhangMallat}
Zhang, S. {\&} Mallat, S. (2019)  Wavelet Phase Harmonic Covariance Models of
  Stationary Processes. {\em submitted to Jour. of Pure and Applied Harmonic
  Analysis}.

\end{thebibliography}

\end{document}